\documentclass{egpubl}
\usepackage{pg2025}
\usepackage{pdfpages}

\SpecialIssuePaper         % uncomment for final version of Computer Graphics Forum, special issue
\CGFStandardLicense
\usepackage[T1]{fontenc}
\usepackage{dfadobe} 
\usepackage{url}
\usepackage{amsmath,amssymb}
\usepackage{multirow}
\usepackage{caption}
\usepackage{lipsum} %
\usepackage{xcolor}

\usepackage{cite}   % comment out for biblatex with backend=biber
\BibtexOrBiblatex
\electronicVersion
\PrintedOrElectronic
\usepackage{graphicx}

\usepackage{egweblnk} 
\title{TensoIS: A Step Towards Feed-Forward Tensorial Inverse Subsurface Scattering for Perlin Distributed Heterogeneous Media}

\author[Tiwari et al.]
{\parbox{\textwidth}{\centering Ashish Tiwari$^{1}$\orcid{0000-0002-4462-6086} , Satyam Bhardwaj$^{1}$\orcid{0009-0007-5292-9021}, Yash Bachwana$^{1}$\orcid{0009-0002-5687-6397}, Parag Sarvoday Sahu$^{1}$\orcid{0009-0007-2642-0504}, T.M.Feroz Ali$^{2}$\orcid{0000-0003-4368-5831},\\ Bhargava Chintalapati$^{2}$\orcid{0009-0008-9783-076X}, and Shanmuganathan Raman$^{1}$\orcid{0000-0003-2718-7891}
        }
        \\
{\parbox{\textwidth}{\centering $^1$Indian Institute of Technology Gandhinagar  \hspace{4mm} $^2$Qualcomm
       }
}
}
\begin{document}

% uncomment for using teaser
% \teaser{
%  \includegraphics[width=0.9\linewidth]{eg_new}
%  \centering
%   \caption{New EG Logo}
% \label{fig:teaser}
%}

\maketitle
%-------------------------------------------------------------------------
\begin{abstract}
   Estimating scattering parameters of heterogeneous media from images is a severely under-constrained and challenging problem. Most of the existing approaches model BSSRDF either through an analysis-by-synthesis approach, approximating complex path integrals, or using differentiable volume rendering techniques to account for heterogeneity. However, only a few studies have applied learning-based methods to estimate subsurface scattering parameters, but they assume homogeneous media. Interestingly, no specific distribution is known to us that can explicitly model the heterogeneous scattering parameters in the real world. Notably, procedural noise models such as Perlin and Fractal Perlin noise have been effective in representing intricate heterogeneities of natural, organic, and inorganic surfaces. Leveraging this, we first create HeteroSynth, a synthetic dataset comprising photorealistic images of heterogeneous media whose scattering parameters are modeled using Fractal Perlin noise. Furthermore, we propose Tensorial Inverse Scattering (TensoIS), a learning-based feed-forward framework to estimate these Perlin-distributed heterogeneous scattering parameters from sparse multi-view image observations. Instead of directly predicting the 3D scattering parameter volume, TensoIS uses learnable low-rank tensor components to represent the scattering volume. We evaluate TensoIS on unseen heterogeneous variations over shapes from the HeteroSynth test set, smoke and cloud geometries obtained from open-source realistic volumetric simulations, and some real-world samples to establish its effectiveness for inverse scattering. Overall, this study is an attempt to explore Perlin noise distribution, given the lack of any such well-defined distribution in literature, to potentially model real-world heterogeneous scattering in a feed-forward manner.\\
   Project Page: \url{https://yashbachwana.github.io/TensoIS/}


\end{abstract}  
%-------------------------------------------------------------------------

\begin{figure*}[ht]
    \centering
    \includegraphics[width=\linewidth]{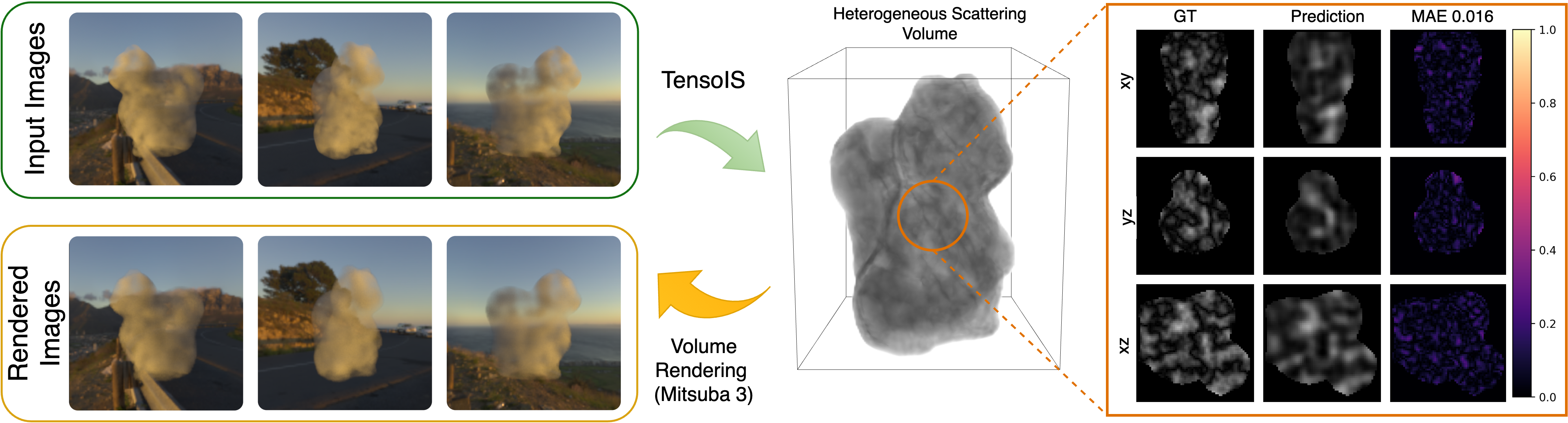}
    \caption{Figure depicts the TensoIS framework designed to obtain the scattering parameters --- extinction coefficient $(\boldsymbol{\sigma_{t}})$ and volumetric albedo $(\boldsymbol{\alpha})$ --- of a heterogeneous medium from six multi-view images. The heterogeneous scattering parameter volume is visualized using MayaVi \cite{ramachandran2011mayavi}. We visualize the $xy$, $yz$, and $zx$ volume slices that maximally cover the cross-sections of the scattering volume. The images rendered (through Mitsuba $3$ \cite{jakob2022mitsuba3}) using predicted scattering parameters are visually similar to the ground-truth images.}
    \label{fig:teaser}
    % \vspace{-0.5cm}
\end{figure*}

\section{Introduction}\label{sec:intro}

Modeling subsurface scattering effects is essential for the realistic rendering of materials like skin, fruits, milk, clouds, smoke, and the atmosphere. Accurately reconstructing scattering parameters in these materials is challenging but essential for photorealistic rendering and understanding optical properties. However, the complex physics of light transport make modeling such media challenging. Their appearance is primarily influenced by subsurface scattering, where light does not travel in straight lines but undergoes multiple interactions within the heterogeneous material characterized by spatially varying optical properties. Estimating heterogeneous subsurface scattering parameters from images, known as \textit{inverse scattering}, is a key research problem in computer vision and graphics and is the primary focus of this paper. \\
\indent While much research has focused on the forward process of photorealistic rendering of heterogeneous media \cite{kallweit2017deep, zhou2008real, bouthors2008interactive, goswami2021survey}, lesser attention has been given to accurately estimating scattering parameters from image observations. Inverse scattering has been explored not only in computer vision and graphics but also in material science, remote sensing, and medical imaging, particularly for CT reconstruction \cite{li2023sparse,auenhammer2024x, ben2024high}. Subsurface scattering is typically modeled using the Bidirectional Scattering-Surface Reflectance Distribution Function (BSSRDF), which describes light transport between surface points \cite{united1977geometrical} and forms the basis for many rendering techniques \cite{chen2004shell, d2011quantized, frisvad2014directional, habel2013photon, vicini2019learned}. However, modeling BSSRDF is challenging due to the complex light paths and multiple bounces within the scattering volume, particularly under spatially varying optical properties (heterogeneous scattering). Many existing inverse scattering methods make simplifying assumptions, focusing either on optically thin (single scattering) \cite{jensen2001practical, narasimhan2006acquiring, deng2024reconstructing} or optically thick (diffusion-based) materials \cite{wang2008modeling}, and fail to cover a wider range of optical densities. Recent advancements in differentiable rendering have combined analysis-by-synthesis with Monte Carlo volume rendering to estimate optical parameters \cite{yang2016inverse, leonard2024image, deng2022reconstructing}. However, they rely on the quality of rendered images to evaluate the accuracy of their parameter estimates. Moreover, these optimization-based methods are slow and prone to local minima (see Figure \ref{fig:abl} (c)). Furthermore, BSSRDF estimation is computationally expensive, requiring complex path integrals and numerous samples. While these methods can produce visually accurate renderings, the estimated parameters may not be physically accurate since there exist multiple material parameter combinations that can produce similar visual appearances \cite{wyman1989similarity,zhao2014high}.\\
\indent In contrast, learning-based feed-forward methods are not only faster but also model complex unknown distributions by leveraging learned priors to provide more consistent and reliable parameter estimates explicitly. The distribution of heterogeneous scattering parameters in the real world is very vivid and highly complex. Additionally, there is no well-established model for sampling heterogeneities corresponding to real-world scattering parameters. A few works have applied a feed-forward learning-based approach \cite{che2020towards, li2023inverse} to estimate subsurface scattering parameters from single-view images by guiding an encoder network to predict parameters that match the ground truth and reproduce the input image. However, they have typically assumed a homogeneous medium. Extracting heterogeneous parameters directly from image observations is even more challenging since (a) subtle variations in scattering parameters may not be visually apparent in the image, and (b) there is no dataset containing images and their corresponding heterogeneous scattering parameter volumes for a feed-forward model to learn from. Interestingly, procedural noise, such as Perlin noise \cite{perlin1985} and Fractal noise (based on Fractal Brownian Motion) \cite{yang2016inverse,perlin1985image}, is known to effectively capture the complex heterogeneities found in natural, organic, and inorganic surfaces. Building on this, we first create a dataset of photorealistic images (using Mitsuba 3 \cite{jakob2022mitsuba3}) of arbitrary 3D shapes where the internal scattering parameters are modeled using Fractal Perlin noise. Through our choices described in Section \ref{sec:dataset}, we generate visually plausible and photorealistic images using Perlin-induced subsurface scattering to mimic real-world appearances closely. Given this dataset, we then propose a learning-based feed-forward framework to model the underlying distribution and estimate the scattering parameters within a bounding volume from sparse multi-view image observations, thus addressing the inverse scattering problem. \\
\indent\textbf{Contributions.} While recent advances have focused on rendering (the forward process), our main goal is to tackle the inverse problem: estimating scattering parameters from image observations (see Figure \ref{fig:teaser}). The following are the key contributions of our work.
\begin{itemize}
    \item We introduce \textit{HeteroSynth}, a synthetic dataset of image-scattering volume pairs. The heterogeneous scattering parameter variations are governed by Fractal Perlin noise and are bounded within arbitrary 3D shapes.
    \item We propose Tensorial Inverse Scattering (TensoIS), a novel learning-based feed-forward framework to estimate spatially varying 3D volume of scattering parameters from a sparse set of multi-view 2D image observations of heterogeneous media bounded within arbitrary shapes.
    \item Instead of directly predicting the bulky 3D volume, TensoIS learns its low-rank tensor components using a set of 2D convolutions.
    % \item To the best of our knowledge, this work presents a dataset containing an image-scattering volume pair for the first time. It represents the first effort to take a feed-forward learning-based approach for estimating subsurface scattering parameters of heterogeneous participating media.
\end{itemize}

\textbf{Note:} This work serves as a step towards modeling \textit{``heterogeneous''} inverse scattering in a feed-forward manner under visible light. Our work is inspired by Che \emph{et al.} \cite{che2020towards}, who were the first to address inverse scattering in a feed-forward manner but only for homogeneous media from a single image. To isolate and better understand the effects of subsurface scattering, we deliberately exclude surface reflection (similar to \cite{che2020towards}), focusing solely on appearance changes arising from subsurface scattering. Surface reflections often largely dominate the object's appearance, and the subtle effect of heterogeneous subsurface scattering lacks sufficient gradients to be discernible by neural networks. With no reflection at the surface, heterogeneous variation within a bounding volume is similar to simulating participating media in the real world.

% The terms heterogeneous volumes or volume tensors would be used interchangeably to mean 3D volumetric grids with spatially varying values. 

% \begin{figure*}
%   \centering
%   \begin{subfigure}{0.68\linewidth}
%     \fbox{\rule{0pt}{2in} \rule{.9\linewidth}{0pt}}
%     \caption{An example of a subfigure.}
%     \label{fig:short-a}
%   \end{subfigure}
%   \hfill
%   \begin{subfigure}{0.28\linewidth}
%     \fbox{\rule{0pt}{2in} \rule{.9\linewidth}{0pt}}
%     \caption{Another example of a subfigure.}
%     \label{fig:short-b}
%   \end{subfigure}
%   \caption{Example of a short caption, which should be centered.}
%   \label{fig:short}
% \end{figure*}

% \vspace{-0.25cm}
\section{Related Works} \label{sec:related_work}

\textbf{Monte Carlo and Diffusion-Based Methods.} Although there are plenty of widely different methods to address the problem at hand, what binds them together is estimating the subsurface characteristics of an object by solving the Radiative Transfer Equation (RTE) \cite{chandrasekhar1960radiative}. It describes the transfer of energy (in our case, light) in the participating media. It had been applied to many areas, including astrophysics and wave propagation, before Blinn \cite{blinn1982light} introduced it in computer graphics. Later, Jakob \emph{et al.} \cite{jakob2010radiative} described a volumetric scattering model to handle scattering media better. Solving RTE has been approached through Monte Carlo (MC) methods and diffusion equations. In MC methods, researchers have proposed different path sampling strategies either by using fixed step distances \cite{blasi1993rendering, blasi1995importance} or sampling cumulative density functions at points over random distances \cite{pattanaik1993computation}. To simplify the light path integrals, several inverse scattering techniques \cite{narasimhan2006acquiring, hawkins2005acquisition, gu2012compressive} relied on single-scattering approximations, which hold good only for optically thin materials. Furthermore, diffusion-based methods introduced a first-order approximation to the radiance \cite{ishimaru1978wave}, proposed a dipole model for subsurface scattering \cite{jensen2001practical, frisvad2014directional, wann2023practical}, and some others introduced finite element methods \cite{stam1995multiple}. Later, \cite{wang2008modeling, li2012transcut} applied the finite element method to cater to heterogeneous media. However, these diffusion-based approaches are only suitable for optically thick media. Gkioulekas \emph{et al.} \cite{gkioulekas2013understanding} and Xiao \emph{et al.} \cite{xiao2012effects} analyze how phase functions, shape, and color influence translucent appearance, highlighting the perceptual and physical factors underlying subsurface scattering. While these studies provide valuable insights into scattering behavior, they focus on analysis and perception rather than inverse reconstruction. \\
\indent\textbf{Inverse Scattering and Differentiable Rendering.} To handle a more general case applicable to different optically dense materials, researchers have resorted to numerical optimization guided by differentiable models of light transport under volumetric scattering  \cite{che2020towards, gkioulekas2016evaluation} by avoiding geometric discontinuities, which itself is challenging and requires computing additional boundary integrals. Some methods used Monte Carlo edge-sampling to sample such discontinuities \cite{li2018differentiable, zhang2022path} and also reduced their variances \cite{zhang2021path, zeltner2021monte}. Recently, Deng \emph{et al.} \cite{deng2022reconstructing} proposed Monte Carlo differentiable rendering for BSSRDF models to handle geometric discontinuities over low sample counts. In addition to these approaches, Yang \cite{yang2016inverse} proposed an iterative scheme to optimize $3$D Simplex noise distribution based on histogram matching of the input and rendered color image. While they obtain the heterogeneous optical parameters through a time-consuming optimization process, they suffer from ambiguity due to similarity relations. The parameters are updated based on the visual appearance of the rendered image, but they do not always guarantee the physical correctness of the optical parameters. Several Neural Radiance Field (NeRF) based methods \cite{levy2023seathru,ramazzina2023scatternerf} and computed tomography based methods \cite{ronen20253d} have also been explored to model participating media such as fog and atmospheric clouds. Furthermore, researchers have developed neural microphysics fields using polarization images \cite{betzer2024nemf} and neural micro-flakes \cite{zhang2023nemf} to perform inverse heterogeneous scattering.\\
\indent\textbf{Deep-learning Based Methods.} Another way to model materials with different scattering characteristics is to use data-driven, deep-learning-based methods. Surprisingly, even with the growing success of these methods, very limited works have explored them to address inverse scattering. Che \emph{et al.} \cite{che2020towards} were the first to consider deep learning for the inverse scattering problem through a single image observation. However, they assumed homogeneous media. Later, Li \emph{et al.} \cite{li2023inverse} extended the earlier method by including surface normal, albedo, and roughness estimation along with homogeneous scattering parameter estimation of translucent objects. Sde-Chen \emph{et al.} \cite{sde20213deepct} also proposed a feed-forward network for atmospheric cloud tomography. However, we focus specifically on Perlin-based real-world heterogeneities and their inference from sparse multi-view images. \\
\indent  In this work, we take a step towards estimating heterogeneous scattering volumes using a $2$D-to-$3$D encoder-decoder framework from a sparse set of multi-view image observations. We believe multi-view images would provide more information for the network to model heterogeneity and handle the underlying ambiguities than just a single image, which is considered sufficient for a homogeneous medium.

% \vspace{-0.25cm}
\begin{figure}[h]
    \centering
    \includegraphics[width=\linewidth]{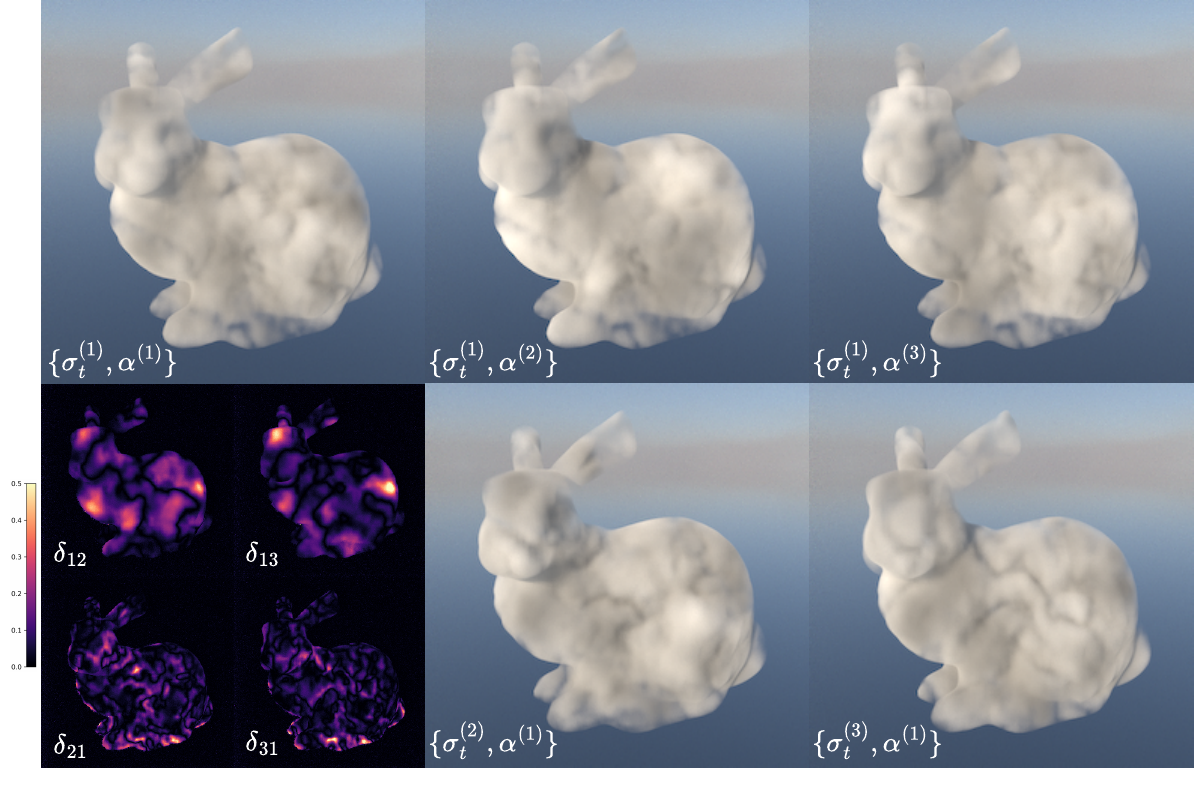}
    \caption{Different visual appearance under subsurface scattering arising due to variation in extinction coefficient ($\sigma_{t}$) and volumetric albedo ($\alpha$). Reference image (top left) and associated difference maps (bottom left).}
    \label{fig:comp}
\end{figure}

\section{Method}
\label{sec:method}

\subsection{Problem Statement}
Given a set of six images of a heterogeneous medium $\mathcal{O}$ acquired from multiple views $\mathcal{I} = \{I_{k}\}_{k=1}^{6}$ under point or environment illumination as an input, we aim to recover the heterogeneous optical parameters that control the scattering of light inside an arbitrarily shaped bounding volume $\mathcal{V}$, a problem popularly known as \textit{heterogeneous inverse scattering}. The scattering of light in a heterogeneous medium is governed by a set of three parameters at each point within the bounding volume.

\begin{itemize}
    \item \underline{Extinction coefficient $(\boldsymbol{\sigma_{t}})$}: is a measure of the attenuation of light as it travels through a medium and quantifies the combined effects of absorption and scattering within the medium i.e. $\boldsymbol{\sigma_{t}} = \boldsymbol{\sigma_{s}} + \boldsymbol{\sigma_{a}} $. It can also be thought of as the probability per unit length that a photon will be scattered or absorbed.\\
    % \vspace{-0.25cm}
    \item \underline{Volumetric albedo $(\boldsymbol{\alpha})$}: is the ratio of the scattering coefficient to the extinction coefficient, i.e.,  $\boldsymbol{\alpha}= \frac{\boldsymbol{\sigma_{s}}}{\boldsymbol{\sigma_{t}}}$. It represents the proportion of light that is scattered (rather than absorbed) as it travels through a medium. A higher albedo indicates that a more significant fraction of the light is scattered, whereas a lower albedo suggests that more light is absorbed.\\ 
    % \vspace{-0.25cm}
    \item \underline{Phase function $(f_{p})$}:  describes the angular distribution of scattered radiation in a particular direction. Particularly, we consider a widely used Henyey Greenstein phase function that models light scattering within the heterogeneous media, like fog, smoke, or biological tissues \cite{toublanc1996henyey}. It is parameterized by ($g$), such that $(g>0)$ indicates forward scattering, $(g<0)$ is backward scattering, and $(g=0)$ indicates isotropic scattering. In this work, we consider isotropic scattering. 

\end{itemize}
Figure \ref{fig:comp} shows how variation of extinction coefficient and volumetric albedo visually affect appearance under subsurface scattering.

\begin{figure*}[t]
    \centering
    \includegraphics[width=\textwidth]{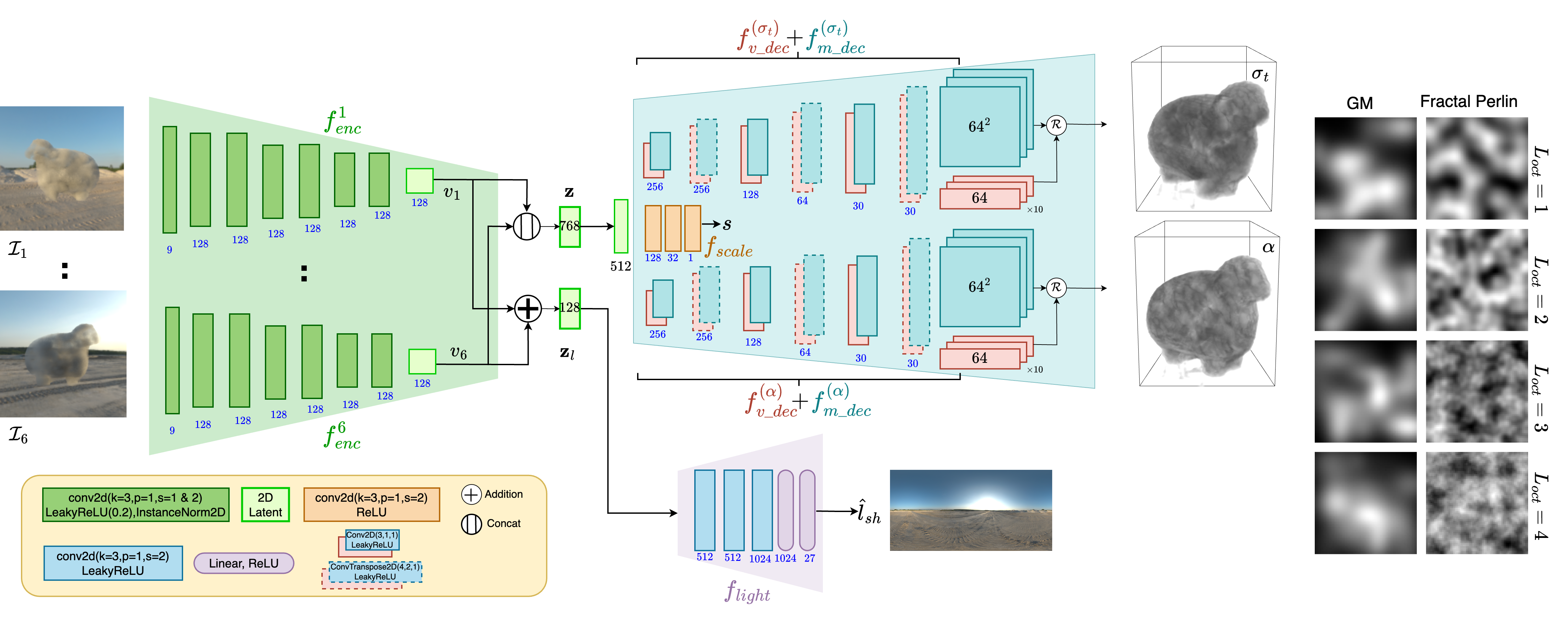}
    \caption{(Left) Architectural details of the proposed TensoIS framework with tensor decomposition-based regression. We use the binary occupancy mask to visualize the heterogeneous variation inside the object. (Right) Textures were generated with Gaussian Mixtures and Fractal Perlin noise distribution for varying octaves. Fractal Perlin noise with $L_{oct}$ = 1 is essentially the Perlin noise.}
    \label{fig:bd}
    % \vspace{-0.25cm}
\end{figure*} 
% \vspace{-0.25cm}
\subsection{TensoIS: Tensorial Inverse Scattering}

We propose a feed-forward deep-learning framework, called Tensorial Inverse Scattering (TensoIS), to estimate the heterogeneous scattering parameter field represented by 3D volume tensors describing extinction coefficient ($\boldsymbol{\sigma_{t}}$) and volumetric albedo ($\boldsymbol{\alpha}$). TensoIS learns low-rank tensor components of the scattering parameter volume, enabling efficient scaling to high-resolution grids, lower memory usage, faster convergence, and effective capture of high-frequency variations in underlying heterogeneities. Although single-image methods may suffice for homogeneous media, heterogeneity is better modeled using multiple views. The proposed auto-encoder-based pipeline is described in Figure \ref{fig:bd} that estimates $3$D scattering volume from $2$D images by using only $2$D convolutions.\\
\indent \textbf{\textit{Image Encoder:}} The image encoder $f_{enc}$ comprises $2$D convolutional feature extractors that obtain features from each of the six multi-view images that are later concatenated to form a latent feature representation $\mathbf{z}$, as per Equation \ref{eq:2}. We use a separate encoder for each view $i$.
\begin{equation}
    \mathbf{f}_i = f_{enc}^{(i)}(\mathcal{I}_{i}) \hspace{2mm} , \hspace{2mm}
    \mathbf{z} = \overset{6}{\underset{i=1}{||}}\mathbf{f}_{i}
    \label{eq:2}
\end{equation}   
For point lighting, the input $\mathcal{I}_{i} = I^{(col\_pt)}_{i} \in \mathbb{R}^{256 \times 256 \times 1}$, is the image under camera co-located point light. However, for environmental lighting $\mathcal{I}_{i} = (I_{i} || I_{i} * M_{i} || M_{i}) \in  \mathbb{R}^{256 \times 256 \times 9}$, where $I_{i} \in  \mathbb{R}^{256 \times 256 \times 3}$ is the image under environment lighting, $M_{i} \in  \mathbb{R}^{256 \times 256 \times 1}$ is the foreground mask, and $||$ represents channel-wise concatenation. The view-specific 2D foreground mask $M_{i}$ is available in the HeteroSynth dataset. Although we use 3-channel (color) RGB images under environmental lighting, for uniform point lighting, a single channel is sufficient to capture shading variations. We found that using more than $6$ views yields only marginal gains, especially in scenes with high-frequency scattering, with diminishing returns beyond $6 - 8$ views. Since each view incurs additional computational cost (processed by a separate encoder), we chose $6$ views as an effective balance between performance and efficiency. \\
\indent\textit{\textbf{Volume Decoder}:} The $3$D volume decoder is also made of $2$D convolutions and estimates low-rank tensor components (vector and matrix components) whose outer product produces the desired $3$D volumes containing extinction coefficients and volumetric albedo. In contrast to other tensor-decomposition-based methods \cite{Chen2022ECCV, jin2023tensoir} that are learned via per-scene optimization (fresh optimization for every scene), ours is a feed-forward approach that produces the scattering volumes of any arbitrary scene during inference. A $3D$ tensor $\mathcal{T} \in \mathbb{R}^{I \times J \times K}$ can be written as a sum of $R$ low-rank components consisting of three vector-matrix pairs, one for each of the orthogonal axis, $X, Y$, and $Z$, such that %as per Equation \ref{eq:3}.  
\begin{align}
\mathcal{T}&=\sum_{r=1}^R \mathbf{v}_r^X \circ \mathbf{M}_r^{Y, Z}+\mathbf{v}_r^Y \circ \mathbf{M}_r^{X, Z}+\mathbf{v}_r^Z \circ \mathbf{M}_r^{X, Y}  
\label{eq:3}
\end{align}
Here, $\circ$ represents the outer product among tensor components, with vectors $\mathbf{v}_{r}^{X} \in \mathbb{R}^{I}$, $\mathbf{v}_{r}^{Y} \in \mathbb{R}^{J}$, $\mathbf{v}_{r}^{Z} \in \mathbb{R}^{K}$, and matrices $\mathbf{M}_{r}^{Y,Z} \in \mathbb{R}^{J \times K}$, $\mathbf{M}_{r}^{X,Z} \in \mathbb{R}^{I \times K}$, $\mathbf{M}_{r}^{X,Y} \in \mathbb{R}^{I \times J}$. We consider four decoder branches $f_{v\_dec}^{(\sigma_{t})}$, $f_{m\_dec}^{(\sigma_{t})}$, $f_{v\_dec}^{(\alpha)}$, and $f_{m\_dec}^{(\alpha)}$, for vector and matrix components of $\boldsymbol{\sigma_{t}}$ and $\boldsymbol{\alpha}$, respectively, as described in Equation \ref{eq:4}. 

\begin{align}
    \mathbf{V}^{(\sigma_{t})} &= f_{v\_dec}^{(\sigma_{t})}(z); \hspace{2mm} \mathcal{M}^{(\sigma_{t})} = f_{m\_dec}^{(\sigma_{t})}(z) \nonumber \\
    \mathbf{V}^{(\alpha)} &= f_{v\_dec}^{(\alpha)}(z); \hspace{2mm} \mathcal{M}^{(\alpha)} = f_{m\_dec}^{(\alpha)}(z)
    \label{eq:4}
\end{align}
Here, ($\mathbf{V}^{(\sigma_{t})}$, $\mathbf{V}^{(\alpha)}$) and ($\mathcal{M}^{(\sigma_{t})}$, $\mathcal{M}^{(\alpha)}$) contains $3R$ vector and $3R$ matrix components along the three orthogonal axes and planes, respectively. We finally take their outer product (as per Equation \ref{eq:3}) to obtain the tensor estimates  $\boldsymbol{\sigma_{t}}$ and $\boldsymbol{\alpha}$. We set $R=10$ components to offer a compression ratio of $47.6\%$ for a tensor of dimension $I=J=K=64$. Each of the decoder branches is a composition of $2$D convolutions. We observed that $2$D convolution for vector components better captured the spatial frequencies than $1$D convolutions or linear layers.\\
\indent\textit{\textbf{Lighting Estimation}:} Additionally, we estimate the Spherical Harmonics (SH) Coefficients to encode the environment lighting via the \textit{lighting estimation module}  such that $\widetilde{l}_{sh} = f_{light}(\mathbf{z}_{l}) \in \mathbb{R}^{3 \times 9} $.
% with a reduced number of parameters of $13.13\text{M}$ compared to $23.44\text{M}$ parameters under direct regression and better performance (see Section \ref{sec:exp}).

\setlength{\tabcolsep}{30.0pt}
\begin{table*}[t]

\centering

\resizebox{\linewidth}{!}{
\begin{tabular}{c|c|c} \hline
\textbf{Attribute} & \textbf{Count} & \textbf{Description} \\ \hline
Shapes & 90 (train) + 13 (test) & 3D triangular meshes \cite{volmap2023} \\      
Voxel grid & $64 \times 64 \times 64$ & Axis-aligned cube of side length 50 cm centered at origin \\ 
$\mathcal{H_{\textsf{fpn}}}$  & 10k (train) + 1k (test) & Fractal Perlin Noise variations with $5$ octaves \\
Optical density (point) & 5  & $ s \in \left[8, 80\right]$\\
Optical density (env.) & 5  & $ s \in \left[30, 130\right]$\\
Views     & 6  & $ \varphi \in \{0, 60, 180, 240, 300\}^{\circ}$, $ \theta = 90^{\circ}$ \\ 
Image size & $256 \times 256$ & $4096$ samples per pixel\\ 
Images (6 views) & $\mathbf{\sim 1.086}$\textbf{M} &  1.08M (train) + 6.63k (test) rendered using Mitsuba 3 \cite{jakob2022mitsuba3}  \\\hline                     
\end{tabular}}
\caption{Statistics of the \textit{HeteroSynth} dataset.}
\label{tab:1}
% \vspace{-0.5cm}
\end{table*}

\subsection{Training Details}
We train the network under point lighting and environment lighting separately. While a point lighting setup is a practical testbed for simulations, heterogeneous scattering media under environmental lighting is profound in the real world. To allow the network to focus only on the points within the scattering media $\mathcal{O}$ in the volumetric grid $\mathcal{V}$, we create a $3$D binary occupancy mask $\mathbf{M_{o}} \in \mathbb{R}^{I \times J \times K}$ (with $I=J=K=64$, in our case). We intend to utilize the network's full capacity to recover scattering parameters within the shape bound. For every bounding shape represented by a triangular mesh, we first compute the signed-distance field (SDF) using \texttt{mesh2sdf} \cite{Wang-2022-dualocnn} at all the grid points in the volume $\mathcal{V}$. We then obtain the binary occupancy mask ($\mathbf{M_{o}}$), $ \forall x \in \mathcal{V}$, such that %as per Equation \ref{eq:7}.
\begin{equation}
     \mathbf{M_{o}}(\mathbf{x}) = \begin{cases}1 & \texttt{SDF}(\mathbf{x}) \leq 0\\0 & \texttt{SDF}(\mathbf{x}) > 0\end{cases}
     \label{eq:7}
\end{equation}
Notably, if the mesh is not already available, we use silhouette-based mesh optimization \cite{nicolet2021large} to deform an icosphere and obtain the bounding shape from six binary image masks that, in turn, are obtained from the Segment Anything model (SAM) \cite{kirillov2023segment}. In fact, while \cite{che2020towards} and \cite{deng2022reconstructing} have optimized a cubic mesh to fit the textured appearance, we deform an icosphere only through silhouettes. 

TensoIS is trained to minimize the masked $L_1$ loss between the ground-truth tensors $\mathcal{T}_{gt}$ and the predicted tensors $\mathcal{T}_{pred}$. Under tensor decomposition-based regression, $\mathcal{T}_{pred}$ is obtained by taking the outer products of the predicted vector-matrix components. With $\mathcal{T} \in \{\boldsymbol{\sigma_{s}}, \boldsymbol{\sigma_{a}}\}$, the volume loss is given by %in Equation \ref{eq:8}.
\begin{equation}
    \mathcal{L}_{vol}(\Theta) = \frac{1}{\underset{i,j,k}{\sum}\mathbf{M_{o}}(i,j,k)}\parallel \mathbf{M_{o}}  \odot (\mathcal{T}_{gt} - \mathcal{F}(\mathcal{I};\Theta))\parallel_{1}
    \label{eq:8}
\end{equation}
Here $\odot$ is the element-wise product, and $\Theta$ is the set of learnable parameters of the TensoIS framework $\mathcal{F}$.  

As discussed, the inherent ambiguity of this problem allows multiple combinations of scattering parameters to produce visually similar appearances \cite{wyman1989similarity, zhao2014high}. However, observing the same heterogeneity under different lighting conditions can help resolve these ambiguities. While slight changes in scattering parameters may not significantly alter the appearance of the medium, changes in lighting can. To exploit this sensitivity, we train the network in a multi-light setup (specifically, two lights at a time). To further enhance the accuracy of scattering volume predictions, we introduce feature regularization ($\mathcal{L}_{reg}$) on the encoder, reinforcing the network’s ability to exploit appearance variations based on lighting differences to yield the same scattering parameters for images of a medium under different lighting. $\mathcal{L}_{reg}$ is defined as:  %per Equation \ref{eq:fm_reg}.
\begin{equation}
    \centering
    \mathcal{L}_{reg} = \frac{1}{N_{z}} (\mathbf{z}_{1} - \mathbf{z}_{2})^{2}
    \label{eq:fm_reg}
\end{equation}
Here, $\mathbf{z}_{1}$ and $\mathbf{z}_{2}$ are features extracted from two differently illuminated images $\mathcal{I}^{(1)}_{i}$ and $\mathcal{I}^{(2)}_{i}$ (say, under two different environment lighting) of the same scattering media and $N_{z}$ is the number of elements in  $\mathbf{z}$. For point lighting, $\mathcal{I}^{(1)}_{i} = I^{(col\_pt)}_{i}$ and $\mathcal{I}^{(2)}_{i} = I^{(left\_pt)}_{i}$ or $I^{(right\_pt)}_{i}$.

% % \setlength{\tabcolsep}{2.0pt}
% \begin{table}[t]

% \centering

% \resizebox{\linewidth}{!}{
% \begin{tabular}{c|c|c} \hline
% \textbf{Attribute} & \textbf{Count} & \textbf{Description} \\ \hline
% Shapes & 90 (train) + 13 (test) & 3D triangular meshes \cite{volmap2023} \\      
% Voxel grid & $64 \times 64 \times 64$ & Axis-aligned cube of side length 50 cm centered at origin \\ 
% $\mathcal{H_{\textsf{fpn}}}$  & 10k (train) + 1k (test) & Fractal Perlin Noise variations with $5$ octaves \\
% Optical density (point) & 5  & $ s \in \left[8, 80\right]$\\
% Optical density (env.) & 5  & $ s \in \left[30, 130\right]$\\
% Views     & 6  & $ \varphi \in \{0, 60, 180, 240, 300\}^{\circ}$, $ \theta = 90^{\circ}$ \\ 
% Image size & $256 \times 256$ & $4096$ samples per pixel\\ 
% Images (6 views) & $\mathbf{\sim 1.086}$\textbf{M} &  1.08M (train) + 6.63k (test) rendered using Mitsuba 3 \cite{jakob2022mitsuba3}  \\\hline                     
% \end{tabular}}
% \caption{Statistics of the \textit{HeteroSynth} dataset.}
% \label{tab:1}
% \vspace{-0.5cm}
% \end{table}

While ($\mathbf{V}^{(\sigma_{t})}$, $\mathbf{V}^{(\alpha)}$) and ($\mathcal{M}^{(\sigma_{t})}$, $\mathcal{M}^{(\alpha)}$) will collectively capture heterogeneous variation, the optical density of the medium can vary by a multiplicative scale factor $s \in \mathbb{R}$, \emph{i.e.}, $s\boldsymbol{\sigma_{t}}$ will correspond to a denser medium with increasing value of $s$. We learn to estimate the optical density $s$ via $f_{scale}$ by minimizing $\mathcal{L}_{scale} = ||s-\widehat{s}||_{2}^{2}$.

Finally, we also include the lighting loss across $N = 6$ number of views, such that %as per Equation \ref{eq:light_loss}.
\begin{equation}
    \centering
    \widehat{l}_{sh} = f_{light}\Bigg(\frac{1}{N}\sum_{i=1}^{N} \mathbf{f}_{i}\Bigg)
    \label{eq:light_loss}
\end{equation}
In the process of estimating environment lighting and reducing $\mathcal{L}_{reg}$, TensoIS forces $\mathbf{z}$ and $\mathbf{z}_{l}$ to capture scattering and lighting-dependent features from input images, respectively.

Thus, the overall objective function is described as follows.
\begin{equation}
\centering
    \mathcal{L}_{total} = \mathcal{L}_{vol} + \mathcal{L}_{scale} + \mathcal{L}_{light} + \lambda\mathcal{L}_{reg}
\end{equation}
Here, $\lambda = 0.1$.

It is important to note that although TensoIS is designed to observe differently lit images of the same heterogeneous medium during training, it does not require multi-lit images during inference. The model is trained with a constant learning rate of $1e-4$ using Adam optimizer with default parameters on \texttt{NVIDIA RTX 4090} and batch size $24$ for over $50$ epochs. The entire dataset has been generated on \texttt{NVIDIA RTX 4090} and \texttt{NVIDIA RTX A5000} GPUs.

\begin{figure*}[ht]
    \centering
    \includegraphics[width=\linewidth]{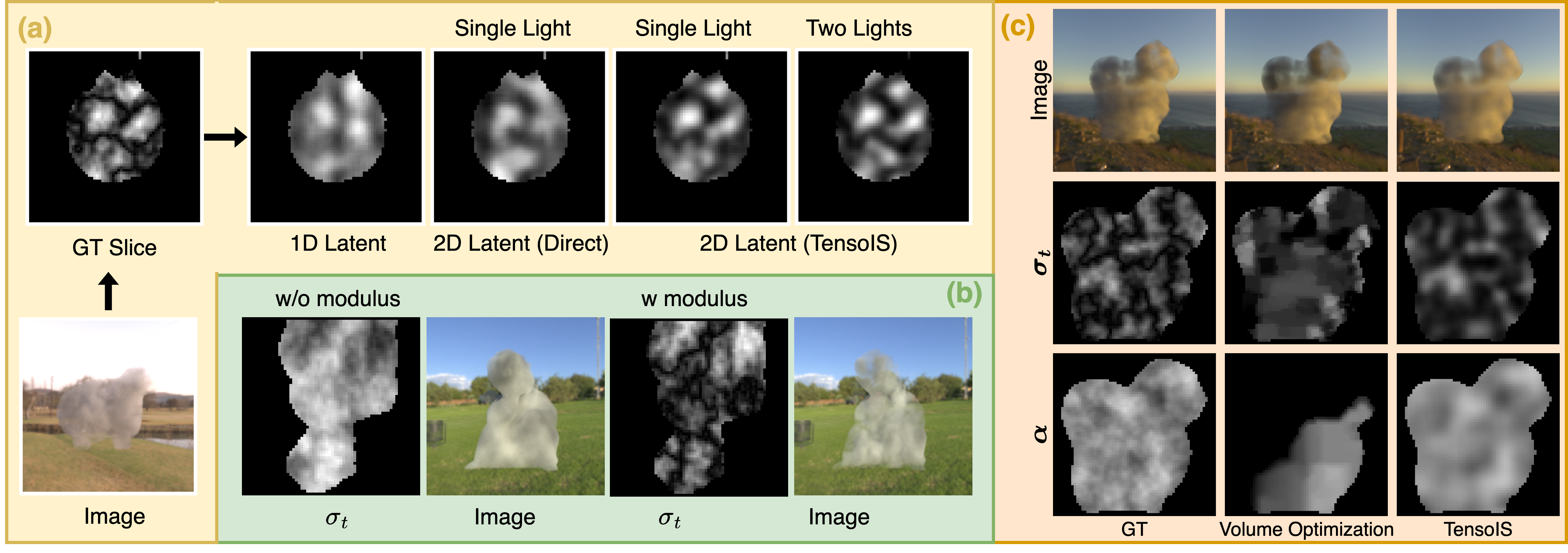}
    \caption{Qualitative effect of (a) varying different design components in TensoIS and (b) introducing modulus operation in $\boldsymbol{\sigma_{t}}$. (c) Qualitative comparison of scattering parameter prediction through TensoIS vs volume optimization. For near-similar-quality rendered images, the estimated parameters are strikingly different.}
    \label{fig:abl}
    %\vspace{-0.5cm}
\end{figure*}
\begin{figure*}[t]
    \centering
    \includegraphics[width=\textwidth]{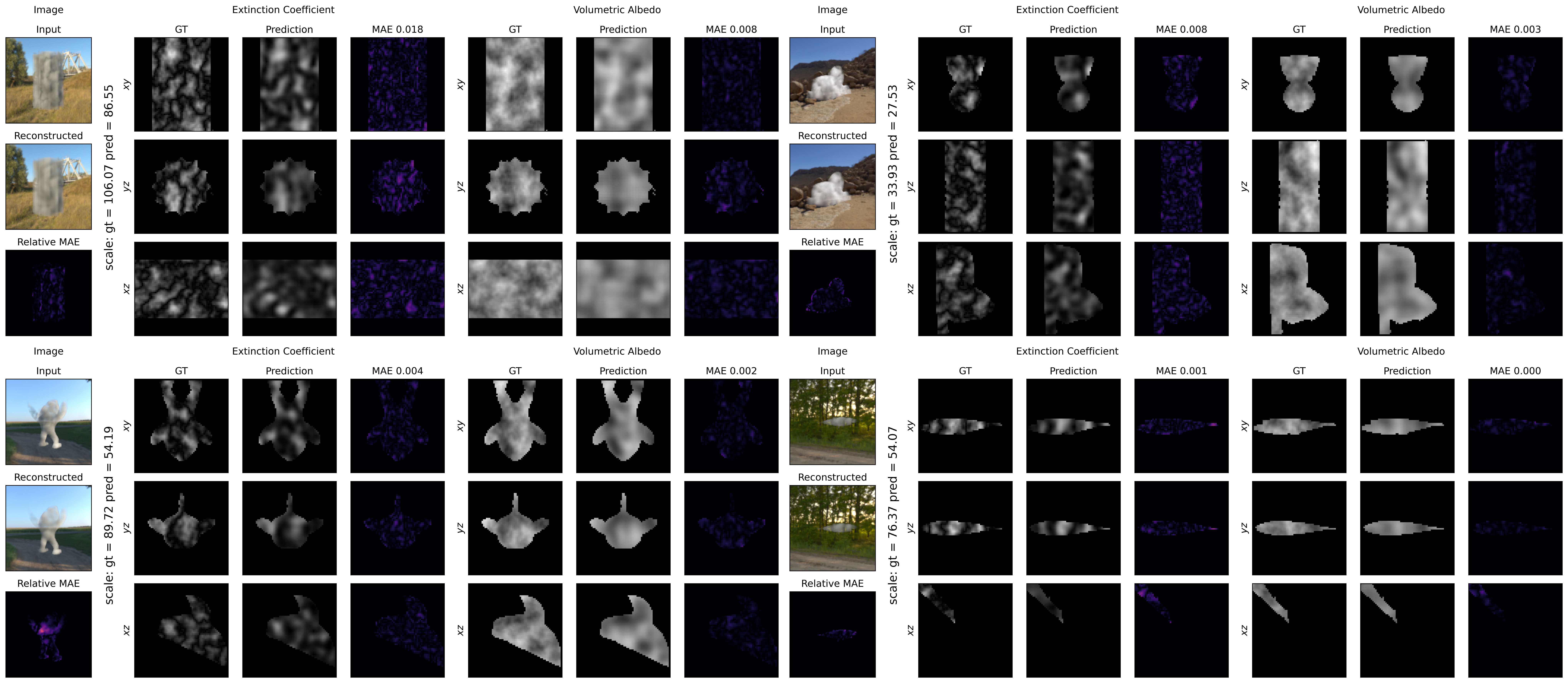}
    \caption{Qualitative results on test set of \textit{Heterosynth} dataset over unseen heterogeneities. Best viewed in pdf with zoom.}
    \label{fig:qual_res}
    % \vspace{-0.25cm}
\end{figure*}

\subsection{Dataset Details}
\label{sec:dataset}

We propose a large-scale synthetic dataset of the arbitrarily shaped heterogeneous medium called \textit{HeteroSynth}. We used $103$ 3D triangular meshes from the VOLMAP dataset \cite{volmap2023} to define the shape of the medium, and the Fractal Perlin Noise model \cite{perlin1985} to generate the heterogeneous scattering parameters ($\boldsymbol{\sigma_{t}}$ and $\boldsymbol{\alpha}$) inside the shape with $g=0$ throughout the medium. However, due to the vast range of variations in Perlin-generated heterogeneities and the need for realistic appearances, we cannot directly use Perlin noise to generate images. Additionally, the inverse scattering problem is highly challenging and ill-posed, especially in the visible spectrum, requiring us to design heterogeneity patterns carefully. We generate 3D fractal Perlin noise values in the range [-1, 1]. For $\boldsymbol{\sigma_{t}}$, we apply a modulus operation to create sharp discontinuities (high-frequency variations, see Figure \ref{fig:abl} (b) for the resulting effect on the image) and vary $\boldsymbol{\alpha}$ within [0.3, 0.95]. We observed that the chosen setting mimics variations in real-world heterogeneous scattering media, such as fog, clouds, or smoke, exhibiting diverse optical densities. Furthermore, for every change in the underlying scattering parameters, these considerations introduce enough photo-realistic visual variations in the images to help the network better model the relation between images and their underlying optical parameters (see Figure \ref{fig:abl} (b)).\\
\indent For a given combination of shape and heterogeneous scattering parameters, we used Mitsuba 3 \cite{jakob2022mitsuba3} to render the medium from six different views under outdoor environment illumination and three-point lights under one-light-at-a-time (OLAT) setup. For each shape in the train set, we rendered $100$ combinations of ($\boldsymbol{\sigma_{t}}$ and $\boldsymbol{\alpha}$) volumes at $5$ different optical densities under point lights, and $50$ different combinations, each under two different environment lighting and $5$ different optical densities. See Table \ref{tab:1} for the complete dataset statistics. We chose to capture six views $60^{\circ}$ apart to ensure full $360^{\circ}$ coverage. Although this may seem like a fixed-view step, the object can always be rotated about the up-axis to obtain a different set of multi-view images. In practice, the network can estimate the scattering parameters from any six views that are $60^{\circ}$ apart and does not need explicit camera parameters. Additional dataset details and results are available in the supplementary material.\\
\indent\textbf{Fractal Perlin Noise.} Perlin noise \cite{perlin1985} is a pseudo-random noise function that is generated by interpolating gradients across a grid and is widely used in computer graphics to create natural-looking textures, terrains, clouds, or smoke. Fractal Perlin noise \cite{perlin1985image} essentially is multiple octaves of Perlin noise at different spatial frequencies and amplitudes. Each octave adds finer detail to the noise, creating a multi-scale effect that resembles fractal patterns found in nature. Figure \ref{fig:bd} shows the textures generated from fractal Perlin noise vs. those by Gaussian mixtures. Mathematically, Fractal Perlin noise $\mathcal{N}_{frac}$ with $L_{oct}$ octaves in 3D, is given by %as per Equation \ref{eq:perlin_noise}.
\begin{equation}
    \centering
    \mathcal{N}_{frac}(\mathbf{x}; N,L_{oct},a)=\frac{1}{Z}\sum_{l=0}^{L_{oct}-1}\frac{1}{2^{l}}\mathcal{N}_{per}(\mathbf{x};N,2^{a + l})
    \label{eq:perlin_noise}
\end{equation}
Here, $\mathcal{N}_{frac}(\mathbf{x}; N,L_{oct},a)$ is the Fractal Perlin noise function evaluated at point $\mathbf{x} \in \mathbb{R}^{3}$ in a regular grid of size $N^3$ for $L_{oct}$ octaves, where $a$ is the starting spatial frequency. $\mathcal{N}_{per}$ is the standard Perlin noise, where the $2^{a+l}$ term corresponds to spatial frequency, which is inversely related to grid spacing as $\frac{N}{2^{a+l}}$. Each octave doubles the frequency, capturing finer details, while the amplitude decreases by a factor of 2. This progressively reduces the impact of higher frequencies for a realistic texture approximation. $Z$ is the normalization factor for the values to lie in the range [-1,1]. In our dataset, we set $L_{oct}=5$ starting from $a=2$ with $N=64$.

\begin{figure*}[t]
    \centering
    % \captionsetup{justification=centering}
    \includegraphics[width=\linewidth]{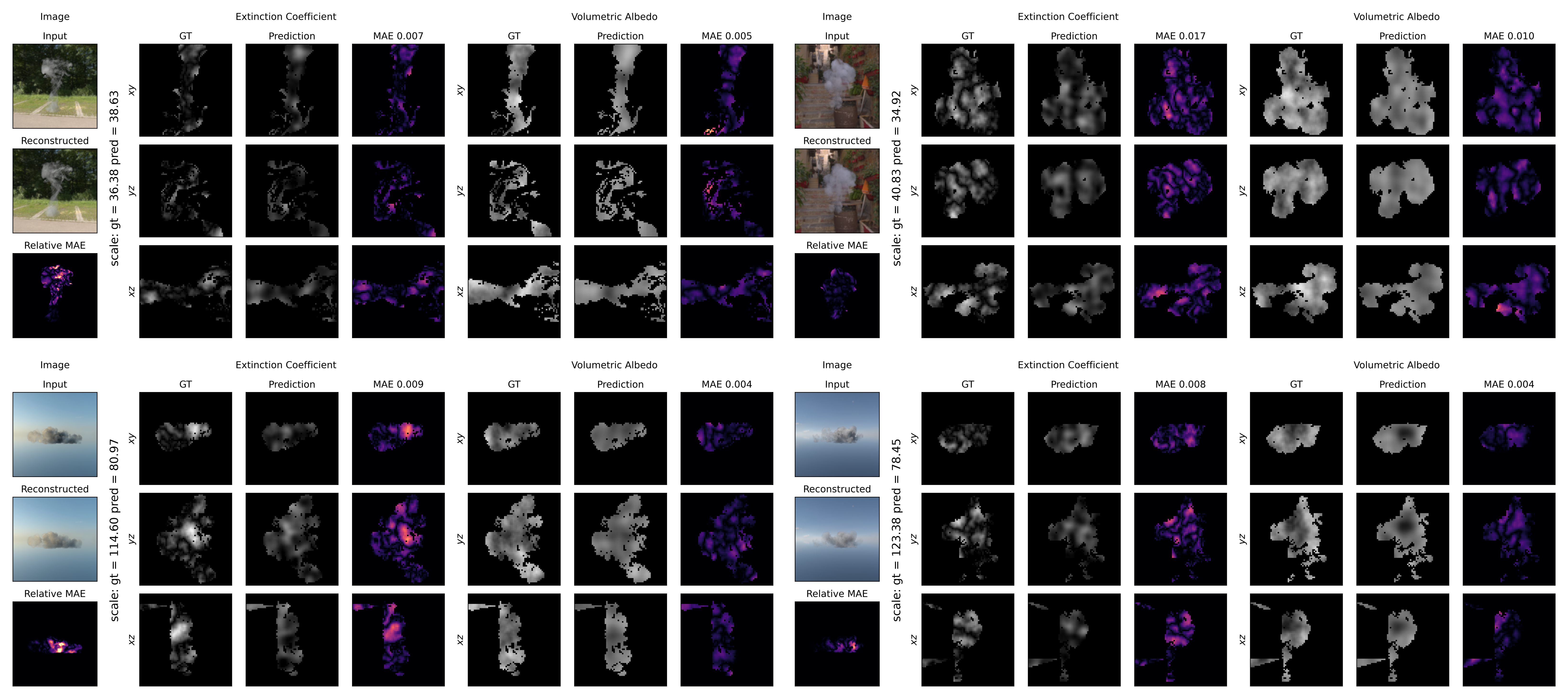}
    \caption{Qualitative results over cloud and smoke geometries. It is best viewed in PDF with Zoom.}
    \label{fig:cloud_smoke}
    
\end{figure*}

\setlength{\tabcolsep}{15pt}
\begin{table*}[h]
\resizebox{\textwidth}{!}{%
\begin{tabular}{c|c|cccc|c|cc}
\hline
\multirow{2}{*}{Design Choices} & \multicolumn{1}{c|}{\multirow{2}{*}{Direct}} & \multicolumn{7}{c}{Tensor Decomposition} \\ \cline{3-9}  
 &  & \multicolumn{1}{c}{1D (\#10)} & \multicolumn{1}{c}{2D (\#1)}  & \multicolumn{1}{c}{2D (\#5)} & \multicolumn{1}{c|}{\textbf{2D (\#10)}} & \multicolumn{1}{c|}{1E (\#10)} & \multicolumn{1}{c}{1D conv(\#10)} & \multicolumn{1}{c}{1D Latent} \\\hline
 $\boldsymbol{\sigma}_{t}$ $(\text{x} 10^{-2})$ & 5.03 & 13.75 & 16.24 & 5.62 & \textbf{3.24}  & 5.77 & 7.14 & 10.73 \\
 $\boldsymbol{\alpha}$ $(\text{x} 10^{-2})$ &\textbf{1.37}& 2.62 & 3.14 & 1.92 & 1.51 & 1.64 & 3.40 & 4.16 \\
 $s$ $(\text{x} 10^{-4})$  & 8.94 & 9.08 & 9.42 & 8.79 & \textbf{8.75}& 9.14 & 8.93 & 10.77 \\
 $\widetilde{l}_{sh}$ $(\text{x} 10^{-4})$ & 2.64 & \textbf{2.61} & 2.79 & 2.68 & 2.62 & 2.73 & 2.65 & 3.03 \\\hline
 $1-$MS-SSIM & 0.0103 & 0.0146 & 0.0189 & 0.0109 & \textbf{0.0098} & 0.0117 & 0.0183 & 0.0471  \\
 MSE & 0.0094 & 0.0101 &0.0122& 0.0083 & \textbf{0.0068} & 0.0082 & 0.0097 & 0.0212  \\\hline
\end{tabular}}
\caption{
We report the Mean Absolute Error (MAE) for the estimated heterogeneous scattering parameters $\boldsymbol{\sigma_t}$ and $\boldsymbol{\alpha}$, as well as the Mean Squared Error (MSE) for the estimated scale and spherical harmonic (SH) coefficients under environment illumination. Additionally, we report $1-$ MS-SSIM and MSE for images rendered using the predicted scattering parameters with Mitsuba 3. Here, $D$ denotes the decoder, $E$ the encoder, 1D conv (in $V$ decoder) (\#N) the number of VM components. Note that 2D (\#10) corresponds to the proposed TensoIS, with each of the other columns representing the effects of alternative design choices within TensoIS.}
\label{table:2}
% \vspace{-0.5cm}
\end{table*}

\begin{figure}[h]
    \centering
    \includegraphics[width=\linewidth]{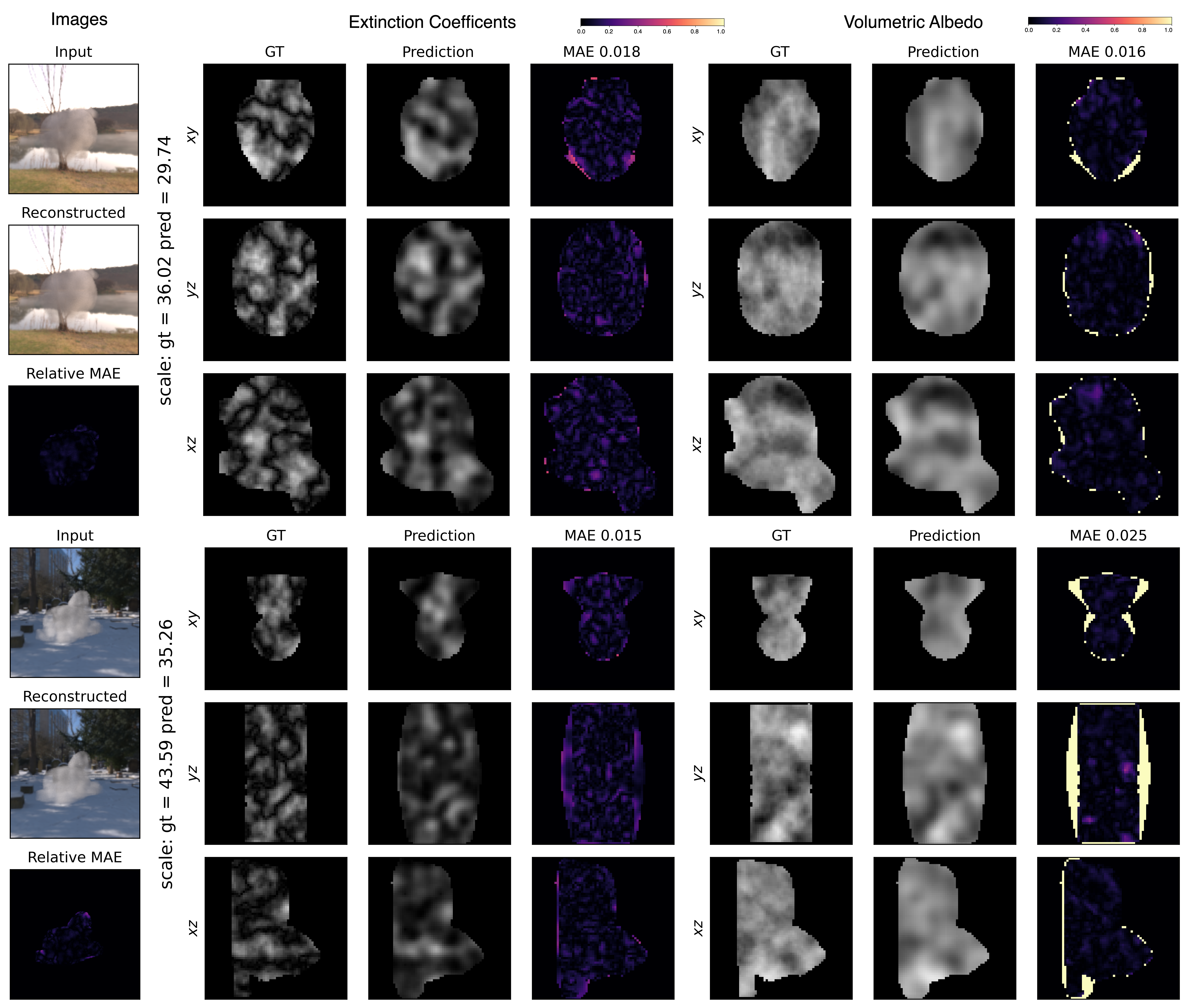}\caption{Effect of using the silhouette-based shape optimization to obtain the bounding mesh (observe the boundaries).}
    \label{fig:shape_optim}
\end{figure}

\section{Experimental Evaluation}
\label{sec:exp}

In this section, we perform an extensive quantitative and qualitative analysis of TensoIS and analyze different architectural design choices.  To visualize the heterogeneous scattering parameters, we show the volume slices (or triplanes, \emph{i.e.} $xy, yz, xz$) planes that maximally contain the cross-section of the object within the underlying 3D grid. Note that the X–Y–Z coordinate convention used for volumetric projections is arbitrarily chosen with consistency across results and does not necessarily align with the image plane convention. We evaluate the TensoIS framework over unseen heterogeneities and report the quality of scattering parameter estimation ($\boldsymbol{\sigma_{t}}$ and $\boldsymbol{\alpha}$) obtained from images under unknown environment lighting. We report MSE between the ground truth and the predicted scattering parameters. Furthermore, we quantify the quality of the images rendered through predicted scattering parameters using MSE and ($1 -$ MS-SSIM), where MS-SSIM is Multi-scale structural similarity. It is observed that the rendered images are very close to the ground truth under unseen heterogeneous parameters.

\subsection{Comparison with existing methods}
This work aims to explicitly estimate Perlin-distributed scattering parameters of heterogeneous media from images under visible light via a feed-forward neural network. The most similar prior work, by Che \emph{et al.} \cite{che2020towards}, estimates scattering parameters for homogeneous translucent objects. Our approach differs in two major ways: (a) we address completely heterogeneous scattering parameters, and (b) we use a regression network rather than an encoder-renderer-based method. Including a renderer, as in prior methods, significantly slows down training due to the computational demands of rendering multi-view images with heterogeneity — a process that is notably faster for single images and homogeneous media. A recent study by Li \emph{et al.} \cite{li2023inverse} extends Che \emph{et al.}’s work \cite{che2020towards} by estimating surface and subsurface scattering parameters from a single image, but still under the assumption of a homogeneous medium.  Given that our approach targets heterogeneous media, direct comparison may not be entirely fair. However, to approximate our model’s performance on homogeneous scattering parameters, we evaluated it on five objects from the test set, with five random draws for $\alpha \in [0.3, 0.95]$ and $\sigma_{t} \in [8, 80]$, chosen to align with ranges used in \cite{che2020towards, li2023inverse}. For each sample rendered under point lighting, we computed the mean and standard deviation of the estimated scattering parameters to assess deviation from the ground truth at each 3D point within the object. Across these samples, we obtained an average MAE (with standard deviation) of $0.4074$ $(\pm 0.0946)$ for $\boldsymbol{\sigma_{t}}$ and $0.1409$ $(\pm 0.0506)$ for $\boldsymbol{\alpha}$. While the mean errors are comparable to those reported in \cite{li2023inverse} on their dataset with single-view image ($0.1590 \pm 0.0023$ and $0.1002 \pm 0.0052$, respectively), the standard deviation is notably higher by orders of $10^{1}$ with higher deviation in $\sigma_{t}$ than in $\alpha$ because of complex variations in $\boldsymbol{\sigma}_{t}$ seen by TensoIS when compared to $\boldsymbol{\alpha}$. Moreover, the network has been trained exclusively on highly heterogeneous data without exposure to homogeneous samples. Predicting a single scalar value for the entire medium (homogeneous) requires an architecture different from predicting per-point values across the entire 3D volume (heterogeneous). While TensoIS has more parameters than \cite{li2023inverse}, we mitigate the higher computational cost to model spatially varying scattering volumes by predicting compact low-rank tensor decompositions instead of full volumetric grids, significantly reducing the memory and obtaining comparable runtime --- as evident upon comparing the time taken for heterogeneous scattering volume reconstruction through TensoIS ($\sim 8.28$ ms) with homogeneous scattering parameter estimation through \cite{li2023inverse} ($\sim 7.80$ ms). Furthermore, while the settings are widely different, we also compare our qualitative results with Deng \emph{et al.} \cite{deng2022reconstructing} in the most comparable setup possible to offer a fair assessment of these methods in the supplementary material.

% \begin{table}[h]
% \resizebox{\linewidth}{!}{%
% \begin{tabular}{l|ccc} \hline
%  & Che et al. \cite{che2020towards} & Li et al \cite{li} & TensoIS \\\hline
% $\sigma_{t}$ & 0.1590 & 0.1590 $\pm$ 0.0023 & 0.4074 $\pm$ 0.0946 \\
% $\alpha$ & 0.1115 & 0.1002 $\pm$ 0.0052 & 0.1409 $\pm$ 0.0051\\\hline
% \end{tabular}%
% }
% \caption{}
% \label{tab:my-table}
% \end{table}

% \vspace{-0.25cm}
\subsection{Ablation Study}

Owing to the above-mentioned reasons, we believe that evaluating different variations of our method and the quality of the images recovered by the estimated scattering parameters would provide a good overall picture of the network performance.\\
\indent\textbf{\textit{Direct Regression vs Low-rank Tensor Decomposition.}}
A straightforward approach to predicting scattering parameters is regressing the 3D parameter volumes using computationally intensive 3D convolutions. In contrast, TensoIS predicts vector-matrix components using lighter 2D convolutions, reducing computational load and accelerating loss convergence by approximately $1.8\times$. TensoIS also avoids the feature-space bottleneck when transferring information from 2D to 3D. This design better captures high-frequency heterogeneities in $\boldsymbol{\sigma_{t}}$, achieving lower MAE in Table \ref{table:2} (row 1). However, for lower-frequency variations (e.g., $\boldsymbol{\alpha}$), both designs perform similarly (Table \ref{table:2} (row 2)).\\
\indent\textbf{\textit{Number of tensor components, decoders, and encoders.}} 
We also examine the effect of varying the number of VM components ($K = 1, 5, 10$) in Table \ref{table:2}. As expected, performance improves with higher $K$, although the gain from $K=5$ to $K=10$ is minimal. We select $K=10$ to achieve a balanced compression ratio of $50\%$, which remains effective even for volume resolutions beyond 64. Given the balance between accuracy and model size, tensor decomposition emerges as the more efficient option. We further observed model degradation when using a single decoder for both $\boldsymbol{\sigma_{t}}$ and $\boldsymbol{\alpha}$, a single encoder for all views, 1D convolution in the V-decoder, or a 1D latent representation of the encoder (both causing heavy loss of spatial information). These choices, individually and collectively, struggled to capture varying degrees of heterogeneity effectively, as shown in Figure \ref{fig:abl} (a).\\
\indent\textit{\textbf{Volume Optimization and Multi-light Training.}}
 Figure \ref{fig:abl} (c) compares TensoIS to volume optimization with total variation (TV) regularization in Pytorch 3D ($\sim 30$ min vs a few ms running TensoIS on \texttt{RTX 4090}). While both approaches produce visually plausible image renderings, the resulting $\boldsymbol{\sigma_{t}}$ and $\boldsymbol{\alpha}$ deviate significantly from the ground truth, with $\boldsymbol{\alpha}$ being overly smooth and washed-out and $\boldsymbol{\sigma_{t}}$ exhibiting blocky artifacts underscoring the ambiguities discussed in Section \ref{sec:intro}. We trained our network with a two-light setup. We applied feature regularization $\mathcal{L}_{reg}$ to ensure consistent latent features for image pairs with identical scattering parameters but different lighting (see Table \ref{table:3}). $\mathcal{L}_{reg}$ enables TensoIS to better disambiguate scattering predictions using multiple lighting conditions, allowing reasonable inference from a single observation. TensoIS performs well with point lighting in the multi-light setup since its variations create more pronounced visual differences than environment lighting. Conversely, environmental lighting performs better in the single-light setup as it introduces illumination from nearly all directions, enhancing the model's capacity to capture finer details (see Table \ref{table:3} ). Unlike optimization-based methods, which are prone to local minima, our data-driven approach with learned priors converges to a more generalizable subspace, offering better performance over unseen Fractal Perlin distributions.

\setlength{\tabcolsep}{20pt}
\begin{table*}[t]
\resizebox{\linewidth}{!}{%
\begin{tabular}{c|ccc|ccc}
\hline

\multirow{2}{*}{Light configuration} & \multicolumn{3}{c|}{Point lighting} &  \multicolumn{3}{c}{Environment lighting}  \\ \cline{2-7}  
 & \multicolumn{1}{c}{Single light} &  \multicolumn{1}{c}{Multi light} & \multicolumn{1}{c|}{w/o $\mathcal{L}_{reg}$} & \multicolumn{1}{c}{Single light} &  \multicolumn{1}{c}{Multi light} & \multicolumn{1}{c}{w/o $\mathcal{L}_{reg}$}  \\ \hline  
 $\boldsymbol{\sigma}_{t}$ $(\text{x} 10^{-2})$  & 5.452 & 2.094 &  4.855 & 4.139 & 3.242 & 4.093\\
 $\boldsymbol{\alpha}$ $(\text{x} 10^{-2})$ & 3.177 & 1.021 & 1.730 & 2.037 & 1.514  &  1.997\\ \hline
 
\end{tabular}}
\caption{Quantitative comparison of TensoIS under single and multi-light training for point and environment lighting along with the effect of feature regularization.}
\label{table:3}
% \vspace{-0.5cm}
\end{table*}

\begin{figure*}[t]
    \centering
    \includegraphics[width=\linewidth]{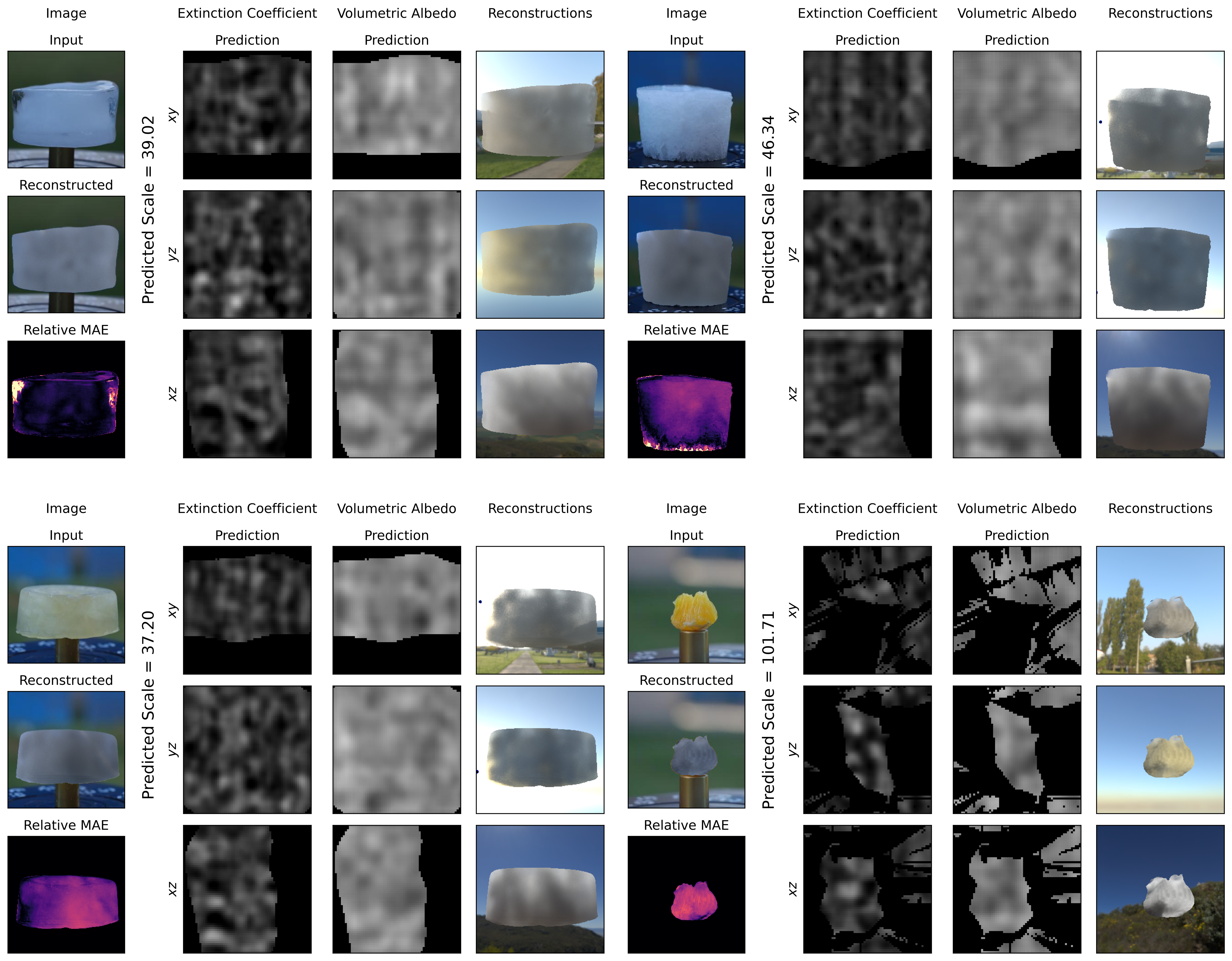}
    \caption{Reconstructions from TensoIS on samples from the real world - ice with tap water, ice with water and soda, ice with water and apple juice, and an orange slice. We also show the appearance of the estimated heterogeneities under different environment lighting rendered using Mitsuba \cite{jakob2022mitsuba3}.}
    \label{fig:real_world}
    % \vspace{-0.5cm}
\end{figure*}

% Table \ref{tab:3} (last column) shows the MSE averaged over 18 different random combinations (6 per distribution) of $\boldsymbol{\sigma_{t}}, \boldsymbol{\alpha}$ across all three distributions, with and without total variation regularization.

\subsection{Qualitative Results}
In Figure \ref{fig:qual_res}, we show extensive qualitative results on the test set of \textit{HeteroSynth} with unseen shapes and heterogeneities and varying optical densities, where the estimated heterogeneous variations are close to the desired ones, along with visually plausible rendering. Figure \ref{fig:cloud_smoke} shows the performance over clouds and smoke geometries. The idea behind showing results on smoke and cloud volumes is solely to demonstrate that Perlin noise distribution can be used to model such media as well. The quality of our results demonstrates that with careful network design and a well-chosen distribution of scattering parameters, it is possible to tackle inverse scattering (even under visible light) by learning effectively from synthetic data.

% \vspace{-0.5cm}
\subsection{Results on Real-world Samples}
At this stage of our study, evaluating real-world data presents several key challenges. (a) Acquiring ground-truth heterogeneous scattering parameters is extremely difficult, making it hard to validate our Perlin distribution-based approximations, (b) matching appearance alone does not guarantee the correctness of the estimated parameters, and (c) our current approach assumes the absence of surface reflection. Although prior works such as \cite{che2020towards, li2023inverse} demonstrate results on a limited set of real-world samples, it is important to note that they also rely solely on appearance-based validation using recovered scattering parameters. Moreover, even though homogeneous parameters are generally easier to obtain than heterogeneous ones, these works do not report the accuracy of the predicted scattering parameters. \\
\indent Following \cite{che2020towards}, we evaluate TensoIS on photographs of heterogeneous translucent objects such as ice formed by plain tap water, mixing water with soda and apple juice, and an orange slice. Our network takes six multi-view images of each object (captured at $60^{\circ}$ intervals) under uncontrolled geometry and natural illumination. To approximate geometry, we modify a base icosphere using silhouette-based differentiable rendering and extract a 3D occupancy mask ($\mathbf{M_{o}}$). The environment illumination is captured using PTGui \cite{pt_gui} to generate a panoramic environment map. Using the predicted scattering parameters and the reconstructed scene setup, we render images using Mitsuba \cite{jakob2022mitsuba3} for comparison. Figure \ref{fig:real_world} shows our results on real-world data. Although the synthesized renderings do not perfectly reproduce the appearance of the original objects, they do capture spatial variations in the estimated heterogeneous scattering parameters. Discrepancies in appearance can be attributed to several other factors, such as inaccuracies in the estimated bounding geometry, the lack of surface reflectance modeling, and the absence of modeling spectral dependence over the scattering parameters. The latter particularly affects our ability to capture color-specific appearance cues since the model predicts a single scale value across all three color channels. We also visualize the predicted scattering parameters under arbitrary environment lighting in Figure \ref{fig:real_world}.\\
\indent Nevertheless, we believe these results mark a promising step toward uncalibrated inverse subsurface scattering from casually captured images. Having explored the feed-forward prediction of heterogeneous scattering parameters, our next direction is to optimize Perlin-distributed scattering volumes (instead of some arbitrary random distribution) directly from real-world observations via differentiable rendering. Once optimized, these volumes can serve as ground truth for training and evaluating TensoIS, thereby enabling a more direct and reliable assessment of parameter prediction quality.

% \begin{figure}[t]
%     \centering
%     \includegraphics[width=\linewidth]{images/collage_2x2.png}
%     \caption{Qualitative results over real-world cloud and smoke models. The results are obtained with known 3D geometry and, thus, the occupancy mask.}
%     \label{fig:cloud_smoke}
%     \vspace{-0.5cm}
% \end{figure}

% \vspace{-0.25cm}
\section{Limitations and Future Work}
The proposed study has certain limitations. Firstly, obtaining geometry from silhouette-based optimization tends to suffer at the object boundaries (both the tri-planes and the rendered images) compared to using the ground truth mesh primarily due to sub-optimal mesh estimation (at the boundaries) in an attempt to roughly approximate the bounding shape as shown in Figure \ref{fig:shape_optim}. Furthermore, the assumption of no surface reflection is limiting for real-world applications. We believe that research advancements in separating surface and subsurface reflections (similar to strategies used in \cite{li2023inverse,li2024deep}) will enhance the significance of this study, creating a robust framework for modeling real-world heterogeneities. Moreover, although images rendered with estimated parameters closely match the actual ones, recovering high-frequency details in volume slices is still limited by the network's ability to extract (high-frequency) scattering information from their (relatively) low-frequency image manifestations, enhancing which would be our future goal. Our goal is also to explore other physics-guided models and neural renderers (instead of relying on Mitsuba) for more advanced parameter estimation and rendering of these heterogeneous properties. Overall, we believe that this study is the starting point for plenty of future research directions on heterogeneous inverse subsurface scattering.
\section{Conclusion} \label{sec:conclusion}

We take a step towards estimating scattering parameters of heterogeneous media from sparse multi-view images by (a) developing and using a synthetic dataset, HeteroSynth, that incorporates heterogeneous optical parameters modeled with procedural Fractal Perlin Noise, (b) leveraging deep learning to model heterogeneous scattering parameters under multi-light setup,  and (c) optimizing low-rank tensor components rather than performing direct tensor regression, a method particularly effective for high-resolution volume grids. We observe that the proposed model generalizes well to unseen Perlin heterogeneities at inference. This study is an attempt to establish Perlin noise distribution to potentially model heterogeneous scattering in the real world, particularly when no well-defined distribution for scattering parameters exists in the literature. While at this stage, we could demonstrate this only by generating photorealistic images mimicking the real world through our design choices, other facets of Perlin distribution warrant more exploration in greater depth and are still an open direction. We hope this work will encourage further research in feed-forward heterogeneous inverse scattering.

\noindent\textbf{Acknowledgment} This work is generously supported by Qualcomm Innovation Fellowship and Jibaben Patel Chair in Artificial Intelligence, IIT Gandhinagar.

% \begin{figure*}[t]
%     \centering
%     % \captionsetup{justification=centering}
%     \includegraphics[width=\linewidth]{images/cloud_smoke_final_small.png}
%     \caption{Qualitative results over cloud and smoke geometries. It is best viewed in PDF with Zoom.}
%     \label{fig:cloud_smoke}
    
% \end{figure*}
% \begin{figure*}[h]
%     \centering
%     \includegraphics[width=0.65\linewidth]{images/fig_shape_optim_2.png}\caption{Effect of using the silhouette-based shape optimization to obtain the bounding mesh (observe the boundaries).}
%     \label{fig:shape_optim}
% \end{figure*}

% \begin{figure*}[h]
%     \centering
%     \includegraphics[width=\linewidth]{images/real_world.png}
%     \caption{Reconstructions from TensoIS on samples from the real world - ice with tap water, ice with water and soda, ice with water and apple juice, and an orange slice. We also show the appearance of the estimated heterogeneities under different environment lighting rendered using Mitsuba \cite{jakob2022mitsuba3}.}
%     \label{fig:real_world}
% \end{figure*}
%-------------------------------------------------------------------------
% bibtex

\bibliographystyle{eg-alpha-doi} 
\bibliography{references}  
\clearpage


\section*{Supplementary material}

# TensoIS: A Step Towards Feed-Forward Tensorial Inverse Subsurface Scattering for Perlin Distributed Heterogeneous Media (Supplementary)

Ashish Tiwari[1] , Satyam Bhardwaj[1] , Yash Bachwana[1] , Parag Sarvoday Sahu[1] , T.M.Feroz Ali[2] ,
Bhargava Chintalapati[2] , and Shanmuganathan Raman[1]

[1]Indian Institute of Technology Gandhinagar    [2]Qualcomm

The supplementary material contains the following components.

- HeteroSynth: Dataset Details
- Real Data Acquisition Setup
- Comparison with Deng *et al.* [DLW*22]
- Additional Qualitative Results

## 1. *HeteroSynth* Dataset

*HeteroSynth:* is a large-scale synthetic dataset rendered using Mitsuba 3 [JSR*22] containing multi-view multi-light HDR images of arbitrarily shaped participating media under point lights or outdoor environment maps, along with corresponding ground truth scattering parameter volumes.

**3D Shapes:** We used 103 genus-0 3D triangular meshes from the VOLMAP dataset [CL23] to define the shape of the medium. 90 were used for training and 13 for testing. Figure 2 shows the shapes in the training and test set of *HeteroSynth*.

**Scattering Parameters:** We consider only isotropic scattering, i.e., $g = 0$ homogeneously throughout the medium, and hence do not estimate the phase function. The extinction coefficient ($\sigma_t$) and volumetric albedo ($\alpha$) are heterogeneously varying inside the medium.

**Volumes:** We used the Fractal Perlin Noise model [Per85] to define the heterogeneous scattering parameters ($\sigma_t$ and $\alpha$) inside the shape. We generated a pool of 10,000 different volumes of size $64 \times 64 \times 64$ to be used for training and a separate pool of 1,000 volumes for testing. We confine the participating medium within an axis-aligned cube of side length 50 cm centered at the origin. We simulate this by resizing and centering the meshes maximally within the cube while preserving the object's aspect ratio, with the voxel grid aligned with the cube. Thus, a single voxel is a cube of side length 7.81 mm. Our goal is to predict $\sigma_t$ and $\alpha$ at each voxel that lies within the medium in this grid from multi-view images.

**Scene:** Each scene consists of a 3D triangular mesh defining the shape of the medium, two volumes controlling the scattering parameters $\sigma_t$ and $\alpha$, a scale factor *s* controlling the optical density, and lighting configuration. For each shape in the train set, we rendered 100 random combinations of $\sigma_t$ and $\alpha$ volumes sampled without replacement from the training pool, at 5 different scales under three different point lights. Each scale was sampled uniformly from $[8, 16], [16, 32], [32, 48], [48, 64], [64, 80]$. For outdoor environment lighting, we rendered 50 random combinations of $\sigma_t$ and $\alpha$ volumes, each under two different environment lighting and 5 different scales sampled uniformly from $[30, 50], [50, 70], [70, 90], [90, 110], [110, 130]$. We considered relatively optically dense media under outdoor environment lighting.

**Camera:** We used the `perspective` camera model with the `hdrfilm` sensor to render the scene at $256 \times 256$ image size with 4096 samples per pixel using the `prbvolpath` integrator. TensoIS works with tone-mapped LDR images as input. The camera was placed 100 cm away from the origin on the z-axis with $45°$ FOV looking at the origin. When rendering under point lights, we considered the camera and lighting setup to be fixed while rotating the medium by $60°$ and rendering under three light positions for a total of 18 image observations per scene. In the main paper, we only used the co-located light for training. Under environment lighting, we consider six cameras placed $60°$ apart symmetrically around the medium in the xz plane at the same distance of 100 cm from the origin, with the y-axis pointing towards the sky.

**Lighting:** The dataset is rendered under two lighting configurations, point lights and outdoor environment maps taken from Poly Haven [HDR15]. Similar to [LNN23], we simulate point lights as radiant spheres 10 cm in diameter with a fixed light intensity of 500 $Wm^{-2}sr^{-1}$. We consider three light positions with the following spherical co-ordinates; co-located $(1.1, 90°, 90°)$, left $(1.1, 0°, 105°)$, and right $(1.1, 180°, 75°)$.

## 2. Real Data Acquisition Setup

A Canon EOS 80D DSLR camera with the standard EF-S18-135mm f/3.5-5.6 IS USM kit lens was used to capture close-up shots of the objects. The objects were placed on a raised platform in an outdoor environment to achieve reduced surface reflection effects. Thereafter, the images were captured from six different views ($60°$ spacing) in auto mode. We additionally captured images with $30°$ spacing as well to use them for mesh estimation.

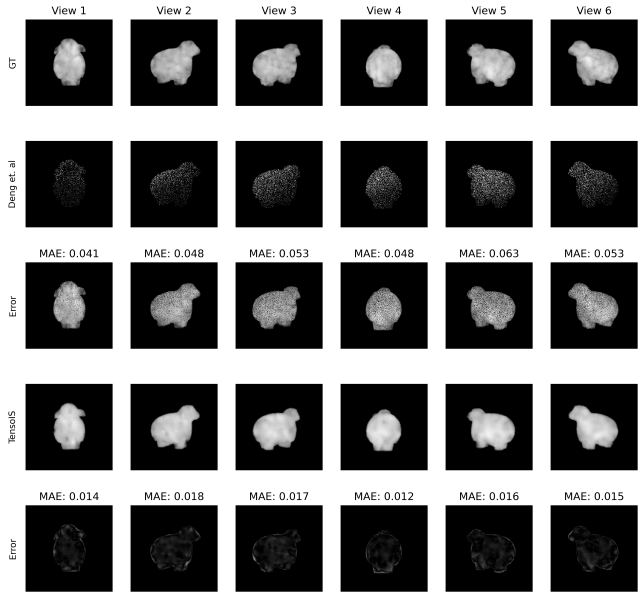

**Figure 1:** *Visual comparison of subsurface scattering parameter estimation of Deng et al. [DLW*22] with TensoIS.*

## 3. Comparison with Deng *et al.* [DLW*22]

Deng *et al.* explicitly models a single-layer BSSRDF via 2D extinction textures through per-scene optimization with 51 input views under point lighting. In contrast, TensoIS operates under environment lighting with only 6 input views and uses a neural network for estimating the full extinction coefficient field. Given these differences in setup, a direct comparison would be fundamentally misaligned. However, for the sake of completeness we use Figure 1 to qualitatively compare the results in the most comparable setup possible to offer a fair assessment of these methods. Specifically, in Figure 1, we evaluate [DLW*22] with 6 views on a sample from the HeteoSynth dataset under point lighting and consider known geometry (*i.e.* we do not optimize the geometry) to keep our focus on scattering parameter estimation. We observe that with 6 input views [DLW*22] is not able to achieve the reconstruction that TensoIS offers.

## 4. Additional Qualitative Results

We also report additional qualitative results on clouds, smoke, and arbitrarily shaped (from *HeteroSynth* test set) participating media under unseen heterogeneities.

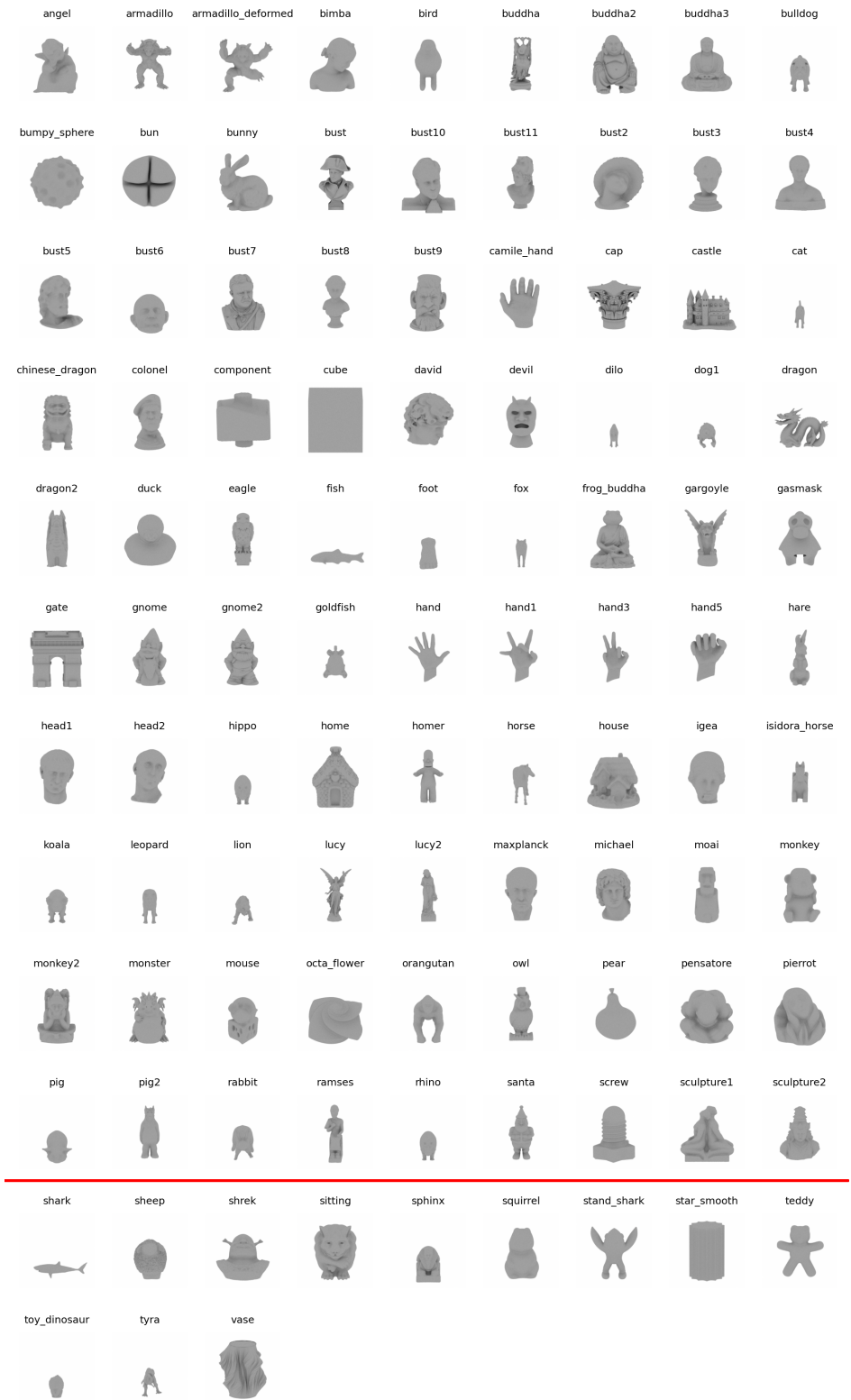

**Figure 2:** *102 genus-0 objects from VOLMAP dataset [CL23] rendered with* `diffuse` *BSDF for illustration. The objects below the red line were used for testing.*

4 of 11 *Tiwari et al. / TensoIS: A Step Towards Feed-Forward Tensorial Inverse Subsurface Scattering for Perlin Distributed Heterogeneous Media(Supplementary)*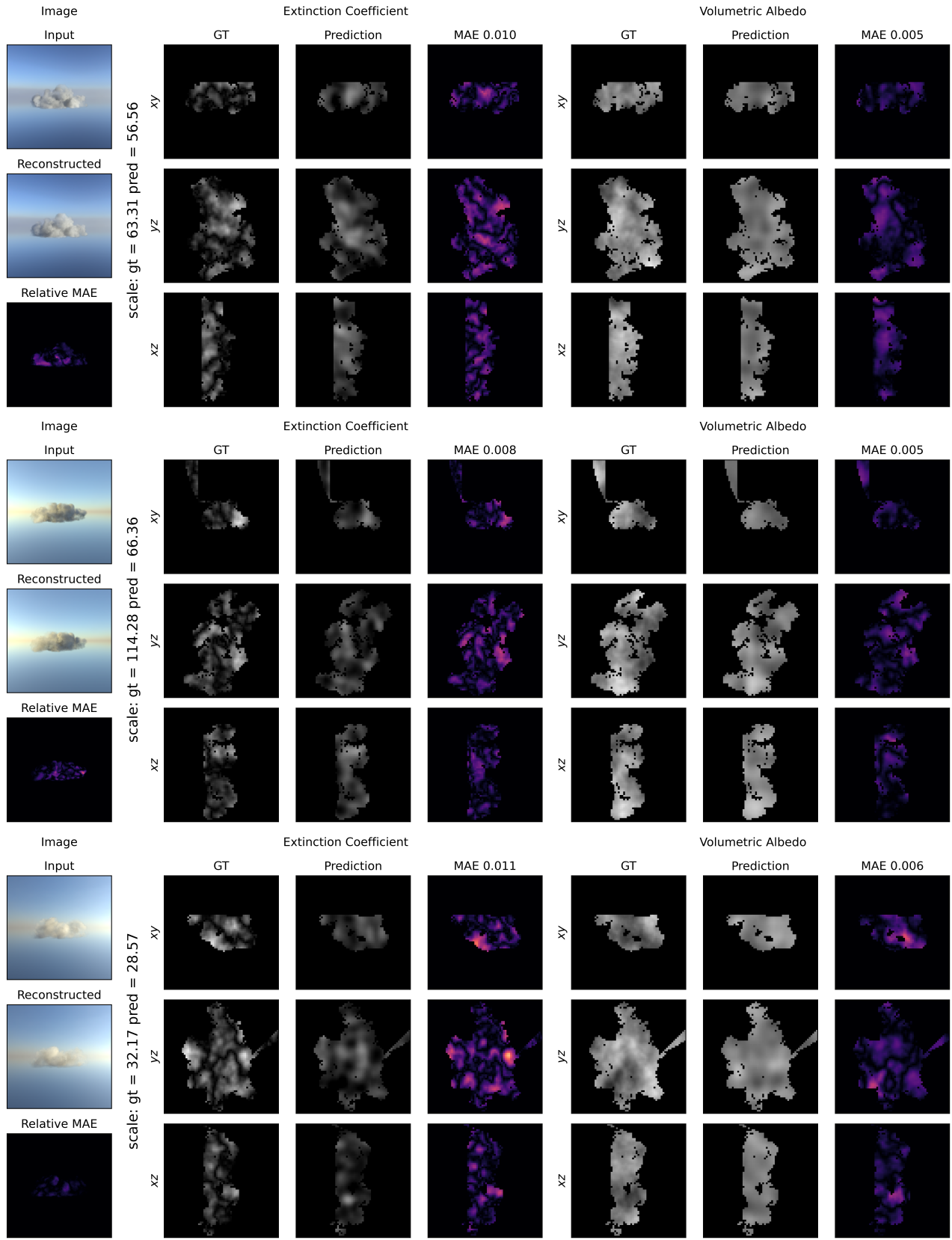

© 2025 Eurographics - The European Association
for Computer Graphics and John Wiley & Sons Ltd.

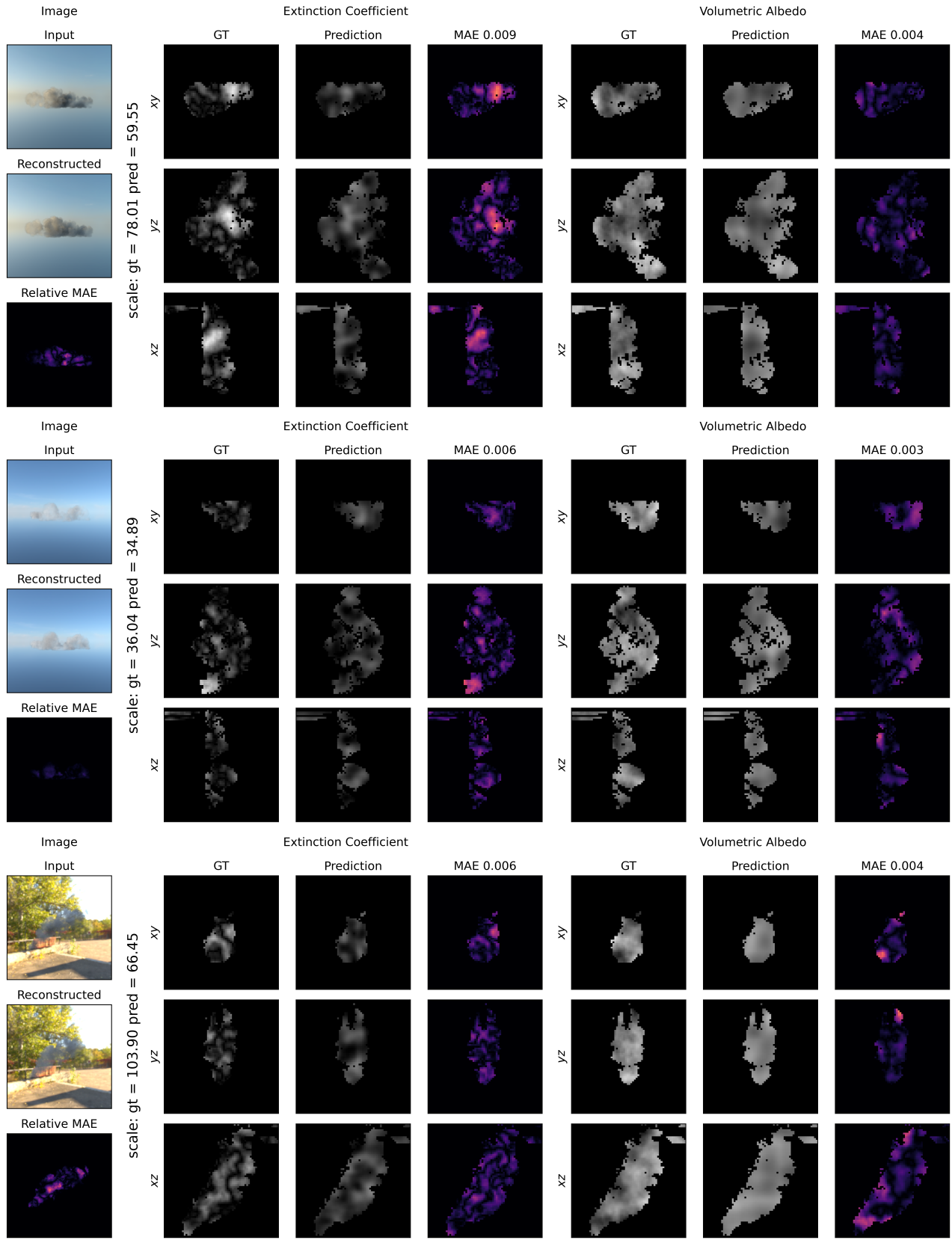

6 of 11 *Tiwari et al. / TensoIS: A Step Towards Feed-Forward Tensorial Inverse Subsurface Scattering for Perlin Distributed Heterogeneous Media(Supplementary)*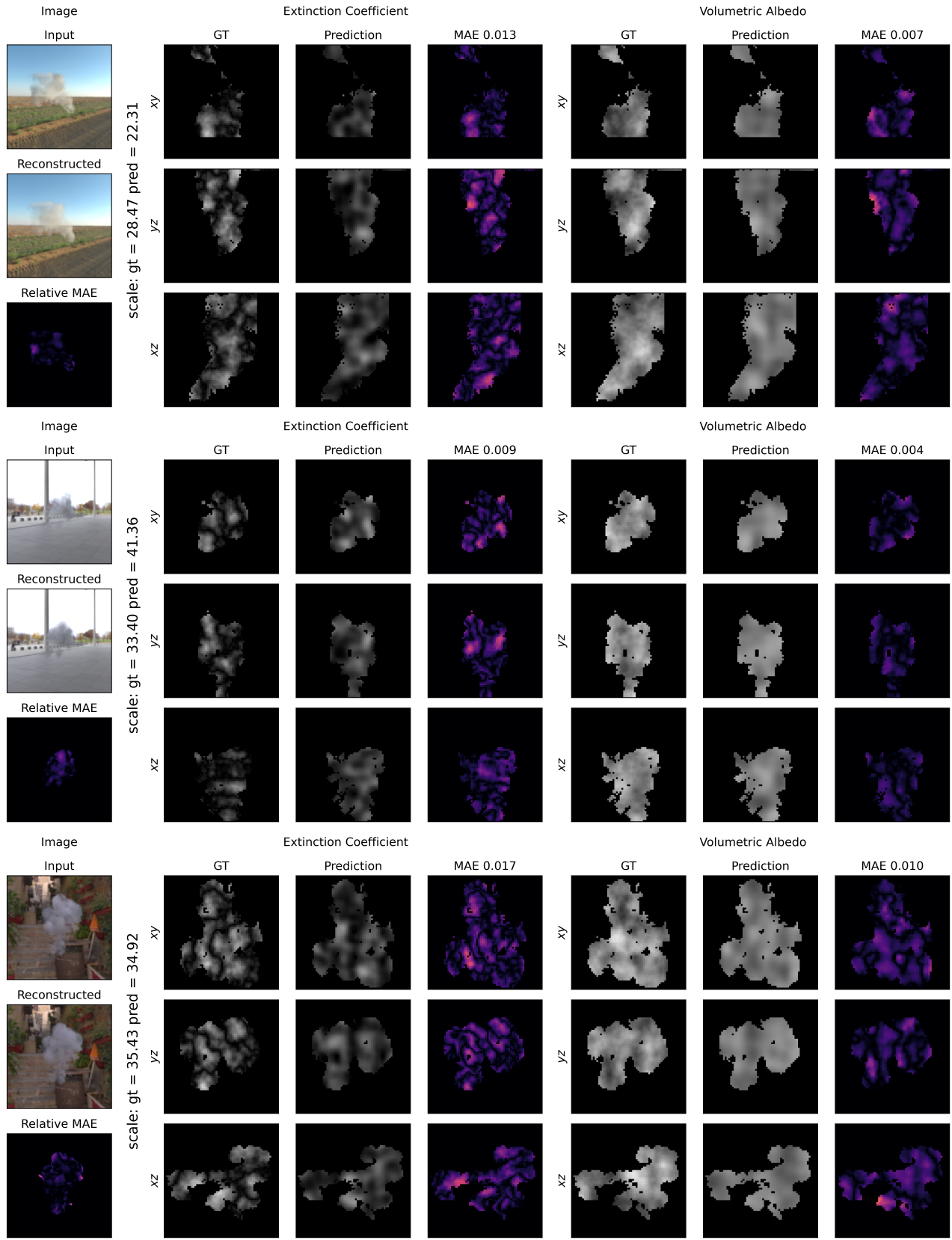

© 2025 Eurographics - The European Association
for Computer Graphics and John Wiley & Sons Ltd.

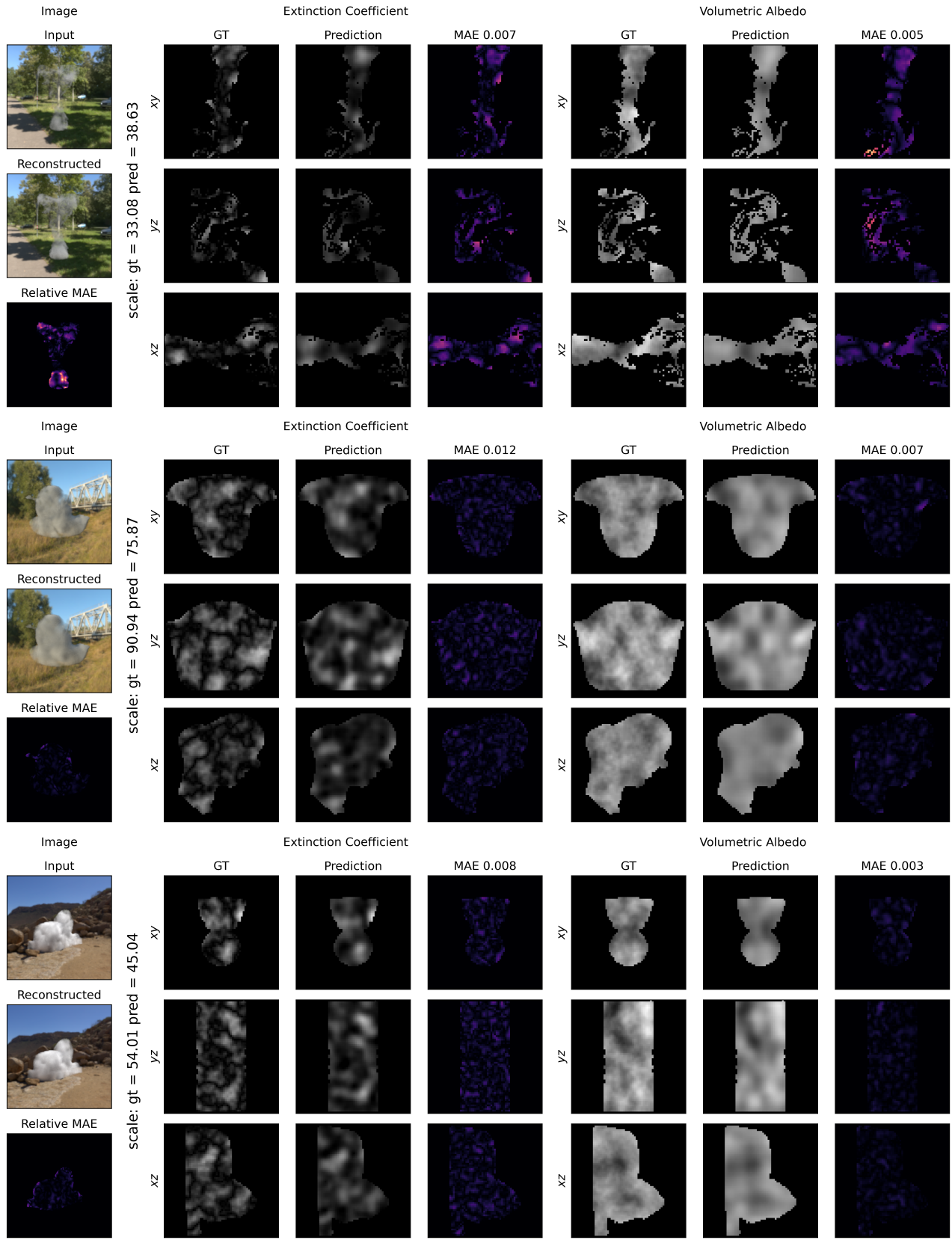

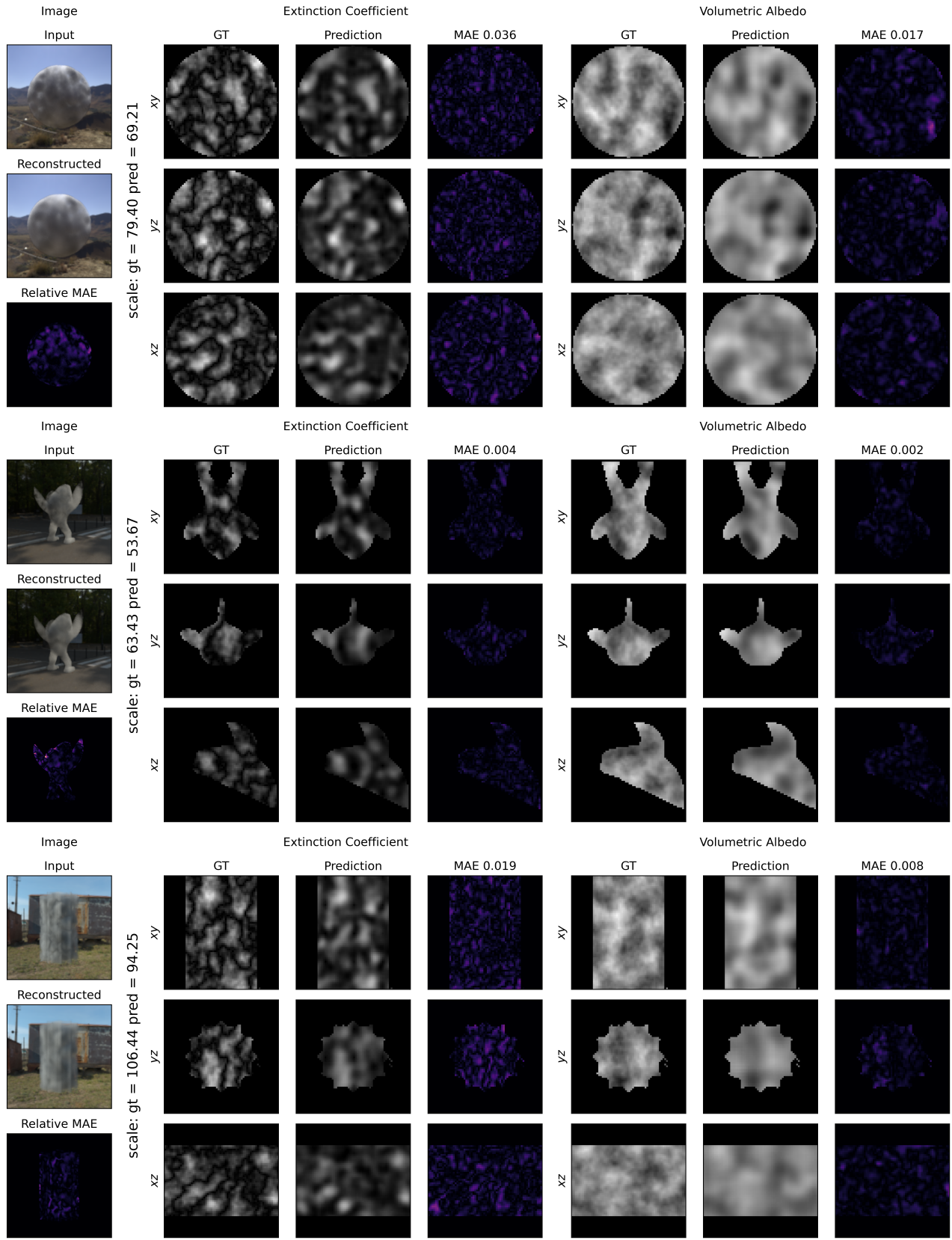

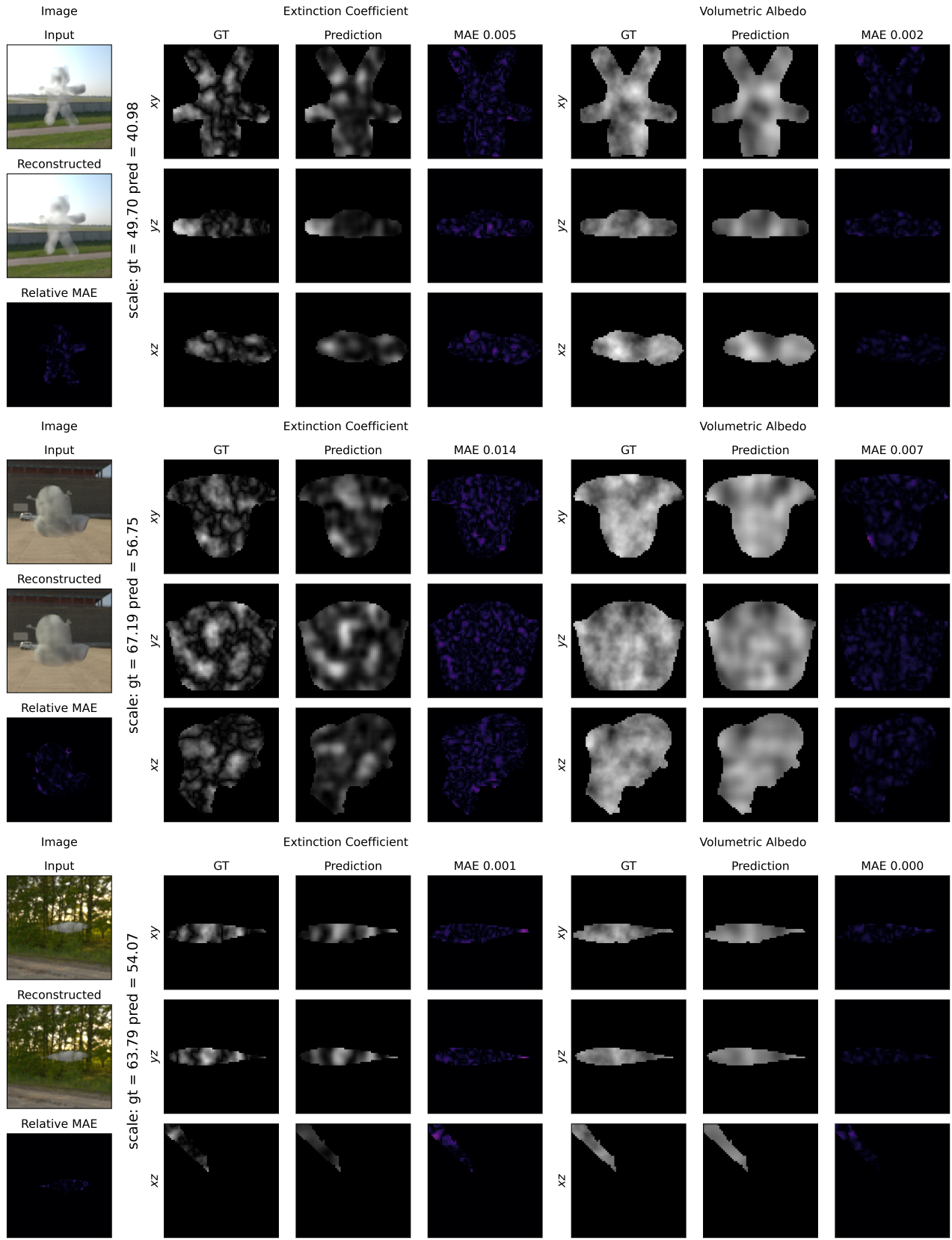

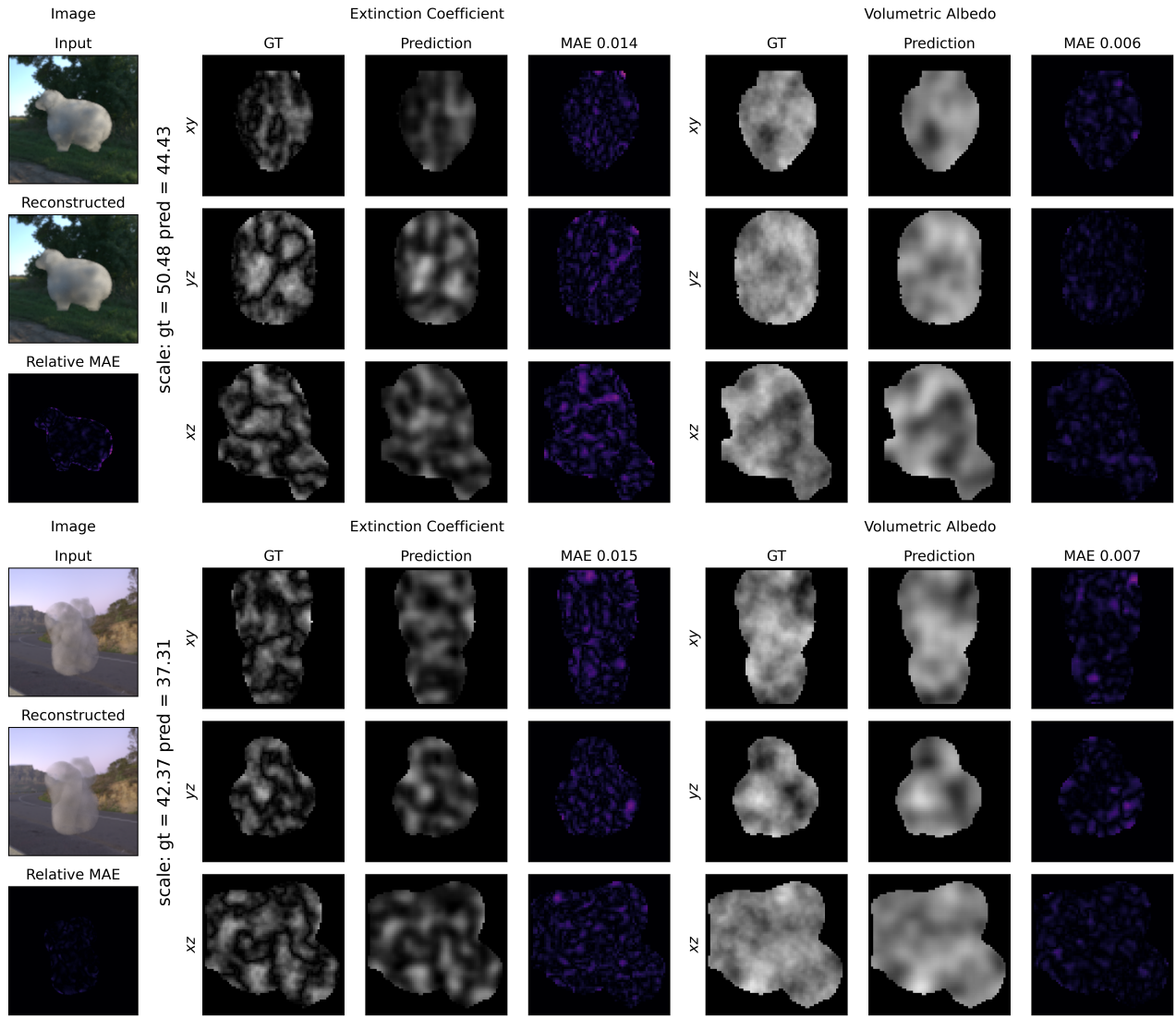



% biblatex with biber
% \printbibliography                

%------------------------------------------------------------------------

\end{document}